\documentclass[journal]{./IEEEtran/IEEEtran}
\IEEEoverridecommandlockouts
\usepackage{cite}
\usepackage{amsmath,amssymb,amsfonts}
\usepackage[dvipsnames,usenames]{color}
\usepackage{array}
\usepackage{algorithmic}
\usepackage{graphicx}
\usepackage{textcomp}
\usepackage{multirow}
\usepackage{float}
\usepackage{mathtools}
\makeatletter
\def\ps@IEEEtitlepagestyle{
	\def\@oddfoot{\mycopyrightnotice}
	\def\@evenfoot{}
}
\def\mycopyrightnotice{
	{ \hspace*{-1cm}\parbox{20cm}{\hrulefill \\ \copyright~2019 IEEE. Personal use of this material is permitted.  Permission from IEEE must be obtained for all other uses, in any current or future media, including reprinting/republishing this material for advertising or promotional purposes, creating new collective works, for resale or redistribution to servers or lists, or reuse of any copyrighted component of this work in other works.}} 
	\gdef\mycopyrightnotice{}
}

\let\old@ps@IEEEtitlepagestyle\ps@IEEEtitlepagestyle
\def\confheader#1{%
	\def\ps@IEEEtitlepagestyle{%
		\old@ps@IEEEtitlepagestyle%
		\def\@oddhead{\strut\hfill#1\hfill\strut}%
		\def\@evenhead{\strut\hfill#1\hfill\strut}%
	}%
	\ps@headings%
}
\makeatother

\confheader{%
	\parbox{20cm}{This work has been accepted for publication in IEEE/OSA Journal of Lightwave Technology, April 2019}
}

\def\BibTeX{{\rm B\kern-.05em{\sc i\kern-.025em b}\kern-.08em
    T\kern-.1667em\lower.7ex\hbox{E}\kern-.125emX}}

\begin{document}

\title{A Variable Rate Fronthaul Scheme for Cloud Radio Access Networks (C-RAN)}

\author{\IEEEauthorblockN{Sandip Das, \IEEEmembership{Member,~IEEE}, and Marco Ruffini, \IEEEmembership{Senior Member,~IEEE}}\\
\IEEEauthorblockA{CONNECT, Trinity College Dublin, Ireland, Email: dassa@tcd.ie, marco.ruffini@scss.tcd.ie}
\vspace*{-3mm}
}

\maketitle

\begin{abstract}
Cloud Radio Access Networks (C-RANs) are considered one of the most promising candidates for implementing 5G mobile communication systems. C-RAN enables centralisation of baseband processing, enabling advanced coordination between base stations, such as coordinated multi-point and inter-cell interference cancellation. In addition, it allows pooling of resources across several cells, providing statistical multiplexing gains of computing resources. However, the link between the remote radio unit and the baseband unit requires high transmission capacity, making the fronthaul link a potential bottleneck for future dense cell deployments.
One of the current solutions to this issue is to compress the fronthaul transmission rate. A second under standardisation is to adopt a different functional split that can reduce the transmission capacity requirement. While this solution decreases the capacity requirements on the transport link, it decentralises some of the computational resources, requiring a more complex remote radio unit (e.g., compared with a CPRI type of solution), whose resources cannot be utilised by other cells when not in use. It is thus expected that in the future, multiple solutions (different functional splits and CPRI) will coexist. 

In this paper, we introduce the concept of Variable Rate Fronthaul (VRF) for C-RAN. This scheme operates on a CPRI type of interface (e.g., one that transmits I/Q data samples) with the novelty of dynamically changing the cell bandwidth, and consequently the fronthaul data rates, depending on the cell load, with the support of a Software Defined Network (SDN) controller. 
This allows for a more efficient transport of C-RAN cells' data over a shared backhaul. We first propose a mathematical analysis of the VRF performance using a queuing theory approach based on the Markov model. We then provide the results of our simulation framework both for validation and in support of the mathematical analysis. Our results show that the proposed VRF scheme provides significantly lower blocking probability over a shared backhaul than standard CPRI.
\end{abstract}

\vspace{0.2cm}

\section{Introduction}
	In the last decade, we have witnessed a massive growth in mobile data connectivity, which is expected to continue steadily for the foreseeable future \cite{CISCOForecastWhitePaper,Loskot2015}. In order to support the demand for higher data rates, seamless mobile connectivity and machine type communication, Fifth Generation (5G) Radio Access Networks (RAN) is being developed with higher cell density and data rates to meet these requirements. Cloud-RAN is a promising framework for 5G, where baseband processing from multiple base stations is centralized (in baseband units - BBUs), while the antenna sites are highly simplified (the remote radio units - RRUs) \cite{C-RAN-Gen-5}. The mobile protocol stack is thus split between the BBU and RRU, and the split point can be chosen among several options (more than eight possible splits are currently being investigated \cite{NGMN_Split_PHY_Arc}). One of these splits, called split-8, and the first one to be deployed in the field through a CPRI interface (Common Public Radio Interface) \cite{CPRI_Standard}, locates all baseband processing in the centralised BBU, leaving only sampling and RF up/down conversion at the RRU. The advantages of the split-8  are those of a simpler RRU, fully centralized resource sharing and the ability to carry out advanced coordinated transmission across multiple cells, such as Coordinated Multi-Point (CoMP) communication \cite{C-RAN-Gen-3}. However, it requires much higher data rates than the legacy distributed RAN (or than a functional split option equal or less than 4, i.e., above the MAC) \cite{C-RAN-Gen-1}.
    
    CPRI supports transport of baseband samples at a fixed rate only and the per Antenna Carrier (AxC) fronthaul capacity demand is higher than 1Gbps (1.2288 Gbps) for a 20MHz LTE RRU \cite{IEEEComMag_CPRI_2016}. When scaled to a future 100MHz bandwidth, 8-channel MIMO systems over three sectors, the required fronthaul capacity reaches 150 Gb/s. In order to overcome these bottlenecks, several solutions have been proposed in recent years, among which compressed CPRI \cite{Compressed_CPRI} and different functional splits \cite{PHY-Split-1} are most popular. The compressed CPRI scheme applies compression in the I/Q baseband samples transported through fronthaul. Though these schemes have the advantage of reducing the fronthaul rate, they nonetheless transport data at a fixed rate, which is independent of the actual cell usage. Another possibility, currently under standardisation is to adopt a different functional split (i.e., below split-8). This reduces the fronthaul capacity demand and enables statistical multiplexing as the fronthaul capacity varies depending on the number of users the RRU is currently serving. However, it is achieved at the expense of a more complex, expensive and power-hungry RRU, as this needs to carry out some part of the physical layer or even MAC processing of the BBU. In addition, any processing resource installed in the RRU remains local to the cell and cannot be shared with other cells. In \cite{stat_Mult_CRAN_1}, the advantage of using centralized processing in C-RAN with a split-8 is analyzed through teletraffic theory and queuing systems. In this work, the authors analyze the improvement of blocking probability in RRU and BBU processing due to C-RAN architecture. Similar work is carried out for C-RAN employing functional split in \cite{stat_Mult_CRAN_2}.
	
    The insight of our work is based on the observation that in future, the progressive densification of mobile cells will dramatically change their traffic patterns. Smaller cells will serve a smaller number of users, leading to much larger statistical fluctuation in cell traffic. Especially when using next generation high-capacity multi-media application, for example, just a few users in one given cell could easily drive it to its maximum rate, while the nearby cells might have no users.
    If such small cells are multiplexed over a shared fronthaul transport system such as a Passive Optical Network (PON), statistical multiplexing can be used to reduce the overall backhaul requirement. 
     However, today statistical multiplexing of C-RAN streams is only possible if functional decomposition is applied with a split-PHY level equal or lower than split-7 \cite{NGMN_Split_PHY_Arc}.
    
	In this paper, we propose a Variable Rate Fronthaul (VRF) scheme that transports the raw baseband I/Q samples as the traditional CPRI transport mechanism (i.e., at split-8). The possibility for commercial TDM-PONs to support CPRI through the use of fixed upstream capacity allocation was experimentally demonstrated in \cite{8385927}. However in this work the upstream capacity is fixed, independently of the actual cell usage. In our scheme instead, a Software Defined Network (SDN) controller monitors the cell usage, and dynamically adapts the cell wireless bandwidth to meet the traffic demand. A reduction in traffic demands will thus trigger a reduction in wireless spectrum usage, which in turn will decrease the I/Q sampling rate and consequently the fronthaul transport rate. The advantage is that the RRU remains simple while restoring statistical multiplexing even for a fronthaul transport system shared between multiple cells (e.g., such as a PON). It should be noted that split-7.1 \cite{NGMN_Split_PHY_Arc}, which puts the Inverse Fast Fourier Transform (IFFT) at the RRU, can also be used with our VRF scheme, since it produces a constant transmission rate (although at a lower rate than split-8). Using both mathematical analysis and event-driven simulation, we evaluated the performance of our proposed scheme for a typical cloud-RAN scenario. The results show that our VRF scheme can achieve significant reduction of traffic congestion in shared fronthaul medium reducing blocking probability of end-user services.
	
	The rest of the paper is organized as follows. Section \ref{sec:SystemModel} provides the system model for the work considered here. Section \ref{sec:Analysis} provides a detailed description of the mathematical analysis of the system based on queuing theory. The details of the simulator framework and the simulation parameters are discussed in Section \ref{sec:SimulationModel}. In Section \ref{sec:Results}, we compare and discuss the results obtained from the analytical method with those obtained through simulation. Finally, in Section \ref{sec:Conclusions} we conclude this article by briefly discussing the outcomes of this work.
	 
\section{System Model} \label{sec:SystemModel}
\begin{figure}[h]
		\centering
		\includegraphics[clip, trim={0in, 0.8in, 0in, 0in }, width=\linewidth]{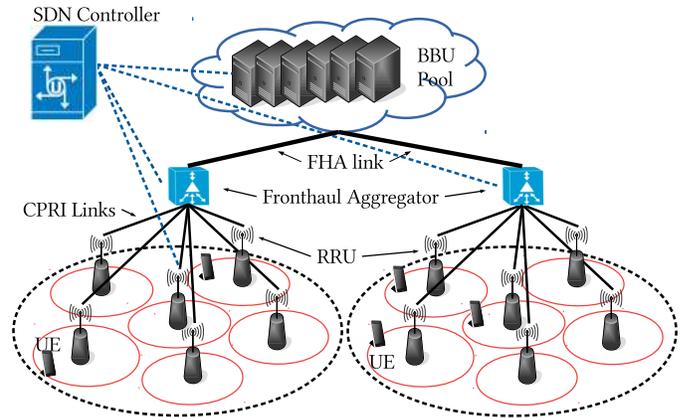}
        \vspace{-0.5cm}
		\caption{\small{System architecture of cloud-RAN}}
		\label{fig:SystemArc}
        \vspace{-0.2cm}
	\end{figure}

	Let us consider the system illustrated in Fig. \ref{fig:SystemArc}. The mobile User Equipments (UEs) are connected to RRUs via an LTE network. Each RRU transmits data to its BBU through CPRI links for centralized processing. We refer to these as fronthaul links. A Fronthaul Aggregator (FHA) aggregates several such links from a cluster of RRUs and creates a Fronthaul Aggregated Link (FAL) which then connects to a centralized BBU pool, forming a tree-like network topology \cite{Tree_topologyBH_Net}. This abstract configuration could be implemented in practice with a Passive Optical Network (PON), by using a power splitter and Optical Line Terminal (OLT) as the fronthaul aggregator, with each RRU having an Optical Networking Unit (ONU) for communicating to the OLT \cite{Ruffni_Mul_conv}. It should be noticed that in general, a shared  backhaul can provide substantial cost savings in dense cell deployments, as it allows to take advantage of statistical multiplexing across base stations. However, statistical multiplexing cannot be exploited by traditional CPRI, which produces constant data rate over the fronthaul link. Our Variable Rate Fronthaul is a solution to this issue: by dynamically changing the wireless bandwidth of each cell and consequently the associated fronthaul rate, it enables both reuse of the wireless spectrum and statistical multiplexing over the shared fronthaul link. 

	In our proposed architecture, the SDN controller interacts with the BBUs to constantly monitor the cell usage of each RRU. As a result, it coordinates operations with the BBU and RRU to dynamically adapt the wireless bandwidth and fronthaul rate. Recently, in \cite{VBF_OFC_paper}, we have experimentally demonstrated the practical feasibility of our variable rate fronthaul concept using a software LTE BBU connected to its RRU (implemented using a USRP board \cite{USRP}) via a fronthaul link operating over a PON. 
	
	In this work, we consider that the UEs are connected to the nearest RRU. Therefore, the cell coverage area per RRU can be estimated analytically using Voronoi diagrams as performed in \cite{stat_Mult_CRAN_2}. However, unlike \cite{stat_Mult_CRAN_2}, where authors assume arbitrary transmission rates depending on user's behavior, and make use of a functional split architecture,  we assume a fully centralized scheme with standard CPRI rates\cite{CPRI_Standard}, listed in the Table \ref{tab:CPRIdataRates}. Each line of the table indicates the wireless bandwidth supported, the size of the Fast Fourier Transform (FFT) required, the number of Physical Resource Blocks (PRBs) supported, the transmission rate at the CPRI interface and the maximum number of users that can be supported. In practice, the number of supported users depends on multiple factors such as traffic type, data rate requirement per user, per user service fairness index, etc. However, in this work, we assume for simplicity that each user is allocated two PRBs (or one Resource Group (RG)) when an application data stream (which we refer to as ``call", following typical queuing theory terminology) is accepted for service. From this, we derive the maximum number of supported users per bandwidth, listed in Table 			\ref{tab:CPRIdataRates}.
		\begin{table}[h]
		\centering
		\caption{Standard CPRI data rates for LTE}
		\label{tab:CPRIdataRates}
		\begin{tabular}{|p{1.75cm}|p{0.8cm}|p{1cm}|p{1.5cm}|p{1.5cm}|}
			\hline
			LTE bandwidth configuration (in MHz)                    & FFT size    & Number of PRBs & CPRI data rate (in Mbps) & Max supported users \\ \hline
			1.25                                                                           & 128      & 6              & 76.8                                                               & 3                   \\ \hline
			2.5                                                                            & 256      & 12             & 153.6                                                              & 6                   \\ \hline
			5                                                                              & 512      & 25             & 307.2                                                              & 12                  \\ \hline
			10                                                                             & 1024     & 50             & 614.4                                                              & 25                  \\ \hline
			15                                                                             & 1536     & 75             & 921.6                                                              & 37                  \\ \hline
			20                                                                             & 2048     & 100            & 1228.8                                                             & 50                  \\ \hline
		\end{tabular}
	\end{table}
	
    In order to explain how the variable rate fronthaul system operates, let us consider a scenario where an RRU is serving already the maximum number of users for the allocated bandwidth. If a new user arrives, which cannot be handled within the current bandwidth, the SDN controller triggers a request to increase the allocated wireless spectrum. As a consequence, the fronthaul CPRI rate also increases to support the higher bandwidth configuration (as listed in table \ref{tab:CPRIdataRates}). Similarly, when a call departs and the number of remaining users can be supported by the next lower bandwidth configuration, both wireless spectrum and fronthaul rate are decreased accordingly.  
    In the following paragraph, we provide a mathematical interpretation of this model.
	
	Let $\phi_{i}$ be the homogeneous Poisson random process with intensity $\Lambda^{i}$, modeling the call arrival process corresponding to the $i^{\text{th}}$ RRU. In this work, we consider the simple case where $\Lambda^{i} = \lambda$, $\forall i \in \{1,2, \dots, N\}$, where $N$ denotes the number of RRUs connected to the same aggregator. Therefore, at any given time instant $t$, the capacity at the FHA link is represented as, \vspace{-0.3cm} 
	\begin{equation}
	\vspace{-0.2cm}
	C(t) = \sum_{i=1}^{N}{\alpha\{u_{i,t}}\}
	\label{eqn:sysModel}
	\end{equation}    
	In (\ref{eqn:sysModel}), $u_{i,t} \in \phi_{i}$ represents the number of users the $i^{\text{th}}$ RRU is serving at time instant $t$. $\alpha:\phi_{i} \to D$, where $D = \{d_1, d_2, \dots , d_M \}$ represents the data rates available for the fronthaul interface. In this work, we also consider the latency introduced by the SDN controller and BBU-RRU system to reconfigure the system on a different bandwidth, which, as shown in \cite{VBF_OFC_paper}, can be estimated in a few hundred milliseconds.
	
	This work aims to find the probability that a new customer call request is blocked, for a given network configuration. Here, we define the call request as the request from an UE to set up a connection with its nearest RRU for transmission and reception of data over one RG. Different network configurations are achieved by varying $N, M, D$ and $\lambda$. Therefore, if $B_C$ represents the capacity of the FHA link, then the steady state blocking probability at the aggregator is given by (\ref{eqn:BlockingProb}) \vspace{-0.2cm}
	\begin{equation}
	\vspace{-0.2cm}
	P_b = \lim\limits_{t \to \infty}\large{P} {\{C(t) \geq B_C \}}
	\label{eqn:BlockingProb}
	\end{equation} 

\section{Theoretical Analysis} \label{sec:Analysis}
	In this section, we provide an analytical solution to the calculation of the blocking probability of a variable rate fronthaul system, based on queuing theory. The process can be divided in two phases. In the first phase, we find the probability for each RRU in a cluster to use  a specific CPRI rate. In the next phase, we find the probability that the aggregator ends up in a blocking state, i.e., where an increase in CPRI rate cannot be supported as the FHA link is already at full capacity. 
	
\begin{figure}
		\includegraphics[width=0.9\linewidth]{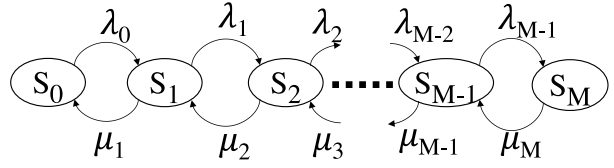}
		\vspace{-0.2cm}
		\caption{\small{State transition diagram for individual RRUs}}
		\vspace{-0.5cm}
		\label{fig:ThresholdQueueMC_0}
	\end{figure}

	
	The queuing analysis for an RRU in a cluster using a given CPRI rate at any time instant requires the use of a threshold-based queuing system. Let $F_i$ represent the forward threshold for an RRU to transition to a higher CPRI rate ($d_i \to d_{i+1}$) and $R_i$ be the reverse threshold for an RRU to transition to a lower CPRI rate ($d_i \to d_{i-1}$). Let $S_i$ denote the state of an RRU using CPRI rate $d_i$ (or in bandwidth configuration $B_i$) with $F_i$ and $R_i$ being the forward and reverse thresholds, respectively. This means that if the RRU is in state $S_i$ and currently serving $F_i$ number of users, then any incoming call request to the RRU will trigger the adoption of the next higher CPRI rate $d_i \to d_{i+1}$ (i.e., the adoption of the next higher bandwidth configuration). Similarly, if the RRU is in state $S_i$, and currently it is serving $R_i$ number of users, then any call departure from the RRU will trigger the adoption of the next lower CPRI rate $d_i \to d_{i-1}$. This state transition is represented in Fig. \ref{fig:ThresholdQueueMC_0}.

	\begin{figure*}
		\includegraphics[width=\linewidth, height = 0.48\linewidth]{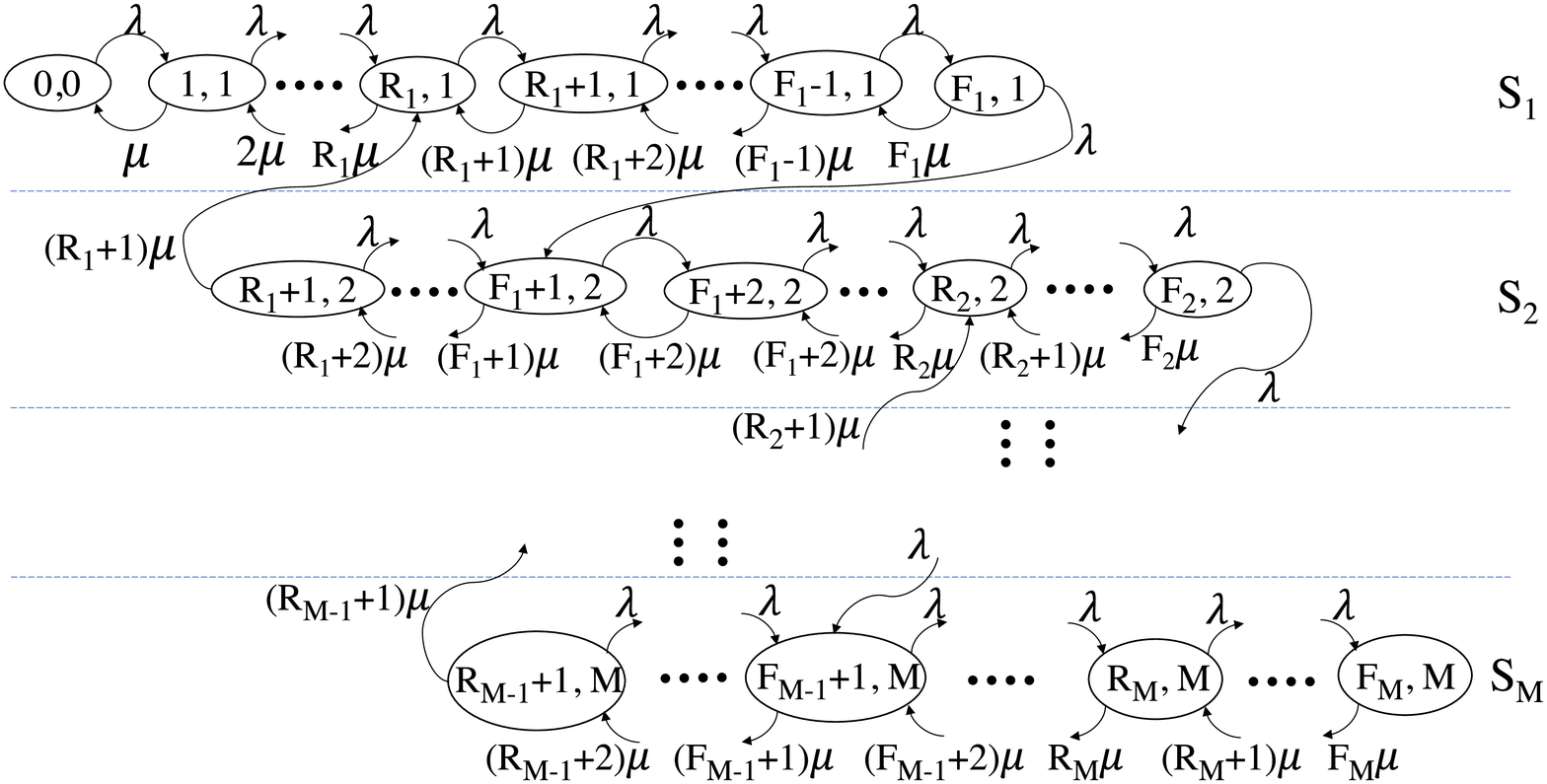}
        \vspace{-0.4cm}
		\caption{\small{State transition diagram of the RRU with partition}}
		\label{fig:ThresholdQueueMC_1}
		\vspace{-0.5cm}
	\end{figure*}

	Our aim in the first phase of the analysis is to find the steady-state probabilities of $S_i$ and the corresponding transition rates $\lambda_{i}$ and $\mu_i$. 
We can clearly notice that the steady-state probability of $S_i$, and the transition rates from $S_i$ to $S_{i+1}$ and  $S_{i-1}$ depend on the sub-states of $S_i$. Thus, in order to provide a solution for this type of system, we use the concept of stochastic complementation technique. This technique has been studied extensively in the field of computer science for adaptive processor allocation in computer servers \cite{StochComp1}, \cite{StochComp2}. In the remainder of this section, we briefly introduce the stochastic complementation technique, before applying it to our model to obtain the steady-state probability and transition rates.
	\vspace{-0.3cm}
	\subsection{Background of Stochastic Complementation}
		The concept of stochastic complementation is based on the classic theory of decomposability of Markov chains which was introduced back in 1977 by P. J. Courtois \cite{DecomposabilityQueueingBook}. The stochastic complementation technique originally introduced by C.D. Meyer in 1989 \cite{Meyer1989StochComp} provides a way of decoupling a Markov chain by means of partitioning. Let $P$ be the transition probability matrix of a discrete space, discrete time Markov chain $\mathcal{M}$ with state space $S$. Let us partition this state space into two disjoint sets $\mathcal{L}$ and $\mathcal{R}$. The one step transition probability matrix of $\mathcal{M}$ becomes: \vspace{-0.1cm}
		\[P= \begin{bmatrix}
		P_{\mathcal{L},\mathcal{L}} & P_{\mathcal{L},\mathcal{R}}\\
		P_{\mathcal{R},\mathcal{L}} & P_{\mathcal{R},\mathcal{R}}\\
		\end{bmatrix}\] 
	 	The steady-state probability vector of $\mathcal{M}$  is $ \pi = [\pi_\mathcal{L}, \pi_\mathcal{R} ] $, where $\pi_\mathcal{L}$ and $\pi_\mathcal{R}$ are the steady-state probability vectors of $\mathcal{L}$ and $\mathcal{R}$, respectively. The stochastic complement of $P_{\mathcal{L},\mathcal{L}}$ which is denoted by $C_{\mathcal{L},\mathcal{L}}$, is given by:
	 	\begin{equation} \label{eqn:StocCompDefn}
	 	C_{\mathcal{L},\mathcal{L}} = P_{\mathcal{L},\mathcal{L}} + P_{\mathcal{L},\mathcal{R}}[I - P_{\mathcal{R},\mathcal{R}}]^{-1}P_{\mathcal{R},\mathcal{L}}
	 	\end{equation}
	 	Let $\pi_{|\mathcal{L}}$ denote the steady-state probability vector corresponding to the states of $C_{\mathcal{L},\mathcal{L}}$. We can write $\pi_{|\mathcal{L}} = \frac{\pi_\mathcal{L}}{\pi_\mathcal{L} . e}$, where $e$ is the column vector with all entries equal to $1$. $\pi_{|\mathcal{L}}$ can be interpreted as the conditional state probabilities of the associated states of the original Markov chain $\mathcal{M}$. 
	 	
	 	We can rewrite (\ref{eqn:StocCompDefn}) as 
	 	\begin{equation}\label{eqn:StocCompDefn1}
	 	C_{\mathcal{L},\mathcal{L}} = P_{\mathcal{L},\mathcal{L}} + \mathop{diag}(P_{\mathcal{L},\mathcal{R}}e)Z
	 	\end{equation}
	 	  In (\ref{eqn:StocCompDefn1}), $\mathop{diag}(v)$ is a diagonal matrix whose $i^{\text{th}}$ diagonal element is the $i^{\text{th}}$ element of vector $v$ and $Z=P^n_{\mathcal{L},\mathcal{R}}[I - P_{\mathcal{R},\mathcal{R}}]  P_{\mathcal{R},\mathcal{L}}$.   $P^n_{\mathcal{L},\mathcal{R}}$ is essentially the matrix $P_{\mathcal{L},\mathcal{R}}$ with all rows normalized. If $r_i$ is the $i^{\text{th}}$ element of $P_{\mathcal{L},\mathcal{R}}e$, and $z_i$ is the $i^{\text{th}}$  row of $Z$, then (\ref{eqn:StocCompDefn1}) can be re-written as:
	 	\begin{equation}\label{eqn:StocCompDefn2}
	 	C_{\mathcal{L},\mathcal{L}}= P_{\mathcal{L},\mathcal{L}} + \begin{bmatrix}
	 								r_1z_1\\
	 								r_2z_2\\
	 								:\\
	 								r_nz_n\\
	 								\end{bmatrix}
	 	\end{equation}
	 	Expression in (\ref{eqn:StocCompDefn2}) can be interpreted as follows. Due to the partitioning of the original Markov process to the sub-processes $\mathcal{L}$ and $\mathcal{R}$, any transition from $\mathcal{L}$ to $\mathcal{R}$ in the original Markov process becomes a transition to some states in $\mathcal{L}$ instead (i.e., it folds back to itself). This process is well known as decoupling of Markov chain. Finding $Z$ can be computationally intensive, although some special cases exist where $Z$ can be easily computed. The following describes one such special case, which we use in our analysis.
	 	
	 	\emph{Theorem (1)}:  Let $Q$ be the transition rate matrix of a given irreducible Markov process with state space $S$. If we partition the state space into two disjoint sets $\mathcal{L}$ and $\mathcal{R}$, then we can write
	 	 \[Q= \begin{bmatrix}
	 	 Q_{\mathcal{L},\mathcal{L}} & Q_{\mathcal{L},\mathcal{R}}\\
	 	 Q_{R,\mathcal{L}} & Q_{\mathcal{R},\mathcal{R}}\\
	 	 \end{bmatrix}\]
	 	In the representation above, $Q_{j,k}$ is the transition rate sub-matrix corresponding to the transitions from partition-$j$ to partition-$k$. If $Q_{\mathcal{R},\mathcal{L}}$ has all zero entries except for some non-zero entries in the $i^{\text{th}}$ column, then the conditional steady-state probability vector (corresponding to the states in $\mathcal{L}$), given that the system is in partition $\mathcal{L}$, is denoted by $\pi_{|\mathcal{L}}$ and is the solution to the following system of linear equations.
	 	\[\pi_{|\mathcal{L}}[Q_{\mathcal{L},\mathcal{L}} + Q_{\mathcal{L},\mathcal{R}}ee^T_i] =\textbf{0}, \qquad \pi_{|\mathcal{L}}e = 1\]
	 	where $e^T_i$ is a row vector with a $0$ in each component except for a $1$ in the $i^{\text{th}}$ component.
	 	
	 	\emph{Proof}: The proof of this theorem can be obtained by following the arguments of the stochastic complementation. We need to apply the simple transformation between a continuous time Markov chain with rate matrix $Q$ and a discrete time Markov chain with probability matrix $P$ which can be obtained via uniformization \cite{CTMC-DTMC_Transform} as provided in (\ref{eqn:PQtransform}). In this equation, $\zeta = \max\{|q_{ii}|\}$, where $q_{ii}$ is the $i^{\text{th}}$ diagonal element of $Q$. For a more detailed proof of this theorem, the reader can refer to \cite{StochComp1}, \cite{StochComp2}.\vspace{-0.1cm}
	 	\begin{equation}\label{eqn:PQtransform}
        P=I+Q/\zeta
        \vspace{-0.3cm}
        \end{equation}
 	\subsection{Steady-state Probability Analysis of the RRU using Stochastic Complementation} \label{sec:thresholdQueue}
 		In our work, we adopt the stochastic complementation method mentioned above to analyze the probability of each RRU using a particular CPRI rate and subsequently its transition rates. We consider the case where all RRUs belonging to the same cluster are characterized by similar arrival rate $\lambda$ and service rate $\mu$.  When a UE call request is accepted, a user session is created at the BBU by allocating an RG to the corresponding UE, over which it transmits and receives data. From this point, we will refer to this user session as `server', following the terminology used in queuing theory. Therefore, the maximum number of servers `$\mathcal{K}$' per RRU is constant and is determined by the maximum number of RGs in the highest bandwidth configuration. 
        Increments and decrements of the CPRI rates or bandwidth configuration are governed by the forward and reverse threshold vectors $F=[F_1, F_2, \dots , F_{M-1}]$ and $R=[R_1, R_2, \dots , R_{M-1}]$, respectively, where $R_i \leq F_i$. 
	 		
 		We consider a homogeneous threshold-based queuing system with hysteresis. Let us construct a Markov process {$\mathcal{M}$}  with the following state space $\mathcal{S}$ according to our system model.\vspace{-0.2cm}
 		\[ \mathcal{S} = \{(N_u, s) \, | \, N_u \geq 0, (s \, | \, d_s \in D)\}\vspace{-0.1cm}\]
 		Where $N_u$ represents the number of users currently served by the given RRU and $d_s$ is the CPRI rate associated with the current bandwidth configuration. Fig. \ref{fig:ThresholdQueueMC_1} illustrates the transition diagram of the Markov process of each RRU. The horizontal lines in the figure are used to partition the whole state space $\mathcal{M}$ into different CPRI spaces. All states $(i,j)$ in any partition $S_j$ are states where the RRU uses the same CPRI rate $d_j$. Horizontal transitions, within the same space $S_j$ are those where the arrival of a call does not trigger a transition to the next higher CPRI rate. This occurs until the number of users reaches the $F_j$ value, after which a further call arrival triggers the transition to the next CPRI space $S_{j+1}$.
A similar process occurs for the transition to lower rates, with the difference that these occur when the number of users, for a given CPRI state $S_j$, decreases below the value $R_j$.
        
        The transition structure of the Markov process $\mathcal{M}$ can be mathematically expressed as follows:
 		\begin{equation}
 		\label{eqn:stateTransitionRRU}
        \hspace{-0.1cm}
	 		\begin{array}{l l}
		 		(0,0) \to (1,1) 	& \quad  \lambda \\
		 		(i,j) \to (i+1, j) 	   & \quad  \lambda \cdot \mathbf{1} \Big \{ j \! \in \! \{ 1, 2, \dots, M\} \wedge \Big( \! (i \notin \! F)\\
		 					     & \quad \quad \vee \big(  (i=F_z \in F) \wedge (j \neq z)  \big)  \Big)\Big\} \\
		 		(i,j) \to (i+1, j+1) & \quad  \lambda \cdot \mathbf{1} \Big \{ j \in \{ 1, 2, \dots, M\}   \\
	 								 & \quad \quad \wedge \, (i=F_z \in F) \wedge (j = z) \Big \} \\
		 		(i,j) \to (i-1, j)	  & \quad \mu \cdot \mathbf{1}  \Big \{(i \geq 1) \wedge \big( (i,j) \neq (1,1) \big) \wedge \\
	 								 & \quad \quad \big( j \! \in \! \{ 1,\dots, M\} \big) \! \wedge \! \big( (i \! - \! 1) \! \notin \! R \big) \\
	 								 & \quad \quad \vee \Big( \!\! \big( \! i \! - \! 1 \! = \! R_z \! \in \! R \big) \! \wedge \! \big( j \! \neq \! z \! + \! 1 \! \big)  \! \!  \Big) \!  \! \Big\}\\
	   			(i,j) \to (i-1, j-1)  & \quad \mu \cdot \mathbf{1}  \Big\{ \! \big( j \! \in \! \{ 1,\dots, M\} \big)  \! \wedge \! \\
		    						 &  \quad \quad \big( \! (i \! - \! 1) \! = \! R_z \! \in \! R \big) \! \wedge \! \big( j \! = \! z \! + \! 1 \! \big) \! \Big\} \\
	   			(1,1) \to (0,0) 	& \quad  \mu \\
	 		\end{array}	
 		\end{equation}
 		In (\ref{eqn:stateTransitionRRU}), $\mathbf{1}\{x\}$ is an indicator function such that $\mathbf{1}\{x\}=1$ if the condition $x$ is true and $0$ if the condition is false. The operators $\wedge \text{ and } \vee$ represent logical ``AND" and ``OR", respectively. 
        
\begin{figure}[h]
	 		 	\vspace{-0.3cm}
	 			\includegraphics[width=\linewidth]{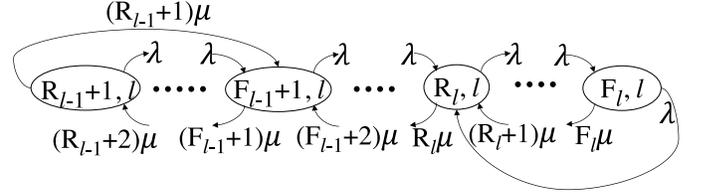}
	 			\vspace{-0.7cm}
	 			\caption{State transition diagram of $\mathcal{S}_l$ for $l \in \{2, \dots, M-1\}$ }
	 			\label{fig:ThresholdQueueMC_2}
	 	\end{figure}

        Let us partition the state space $\mathcal{S}$ into $M$ disjoint sets $\mathcal{S}_l$ $\big( l \in \{1,2, \dots M\}\big)$, where
        \vspace{-0.1cm}
 		\[ \mathcal{S}_l = \{ (i,j) | \quad (i,j) \in \mathcal{S} \,\, and \,\, j=l \}\vspace{-0.1cm}\]
 		Therefore, $\mathcal{S}_l$ represents the state where the RRU is using CPRI rate $d_l$ (or in bandwidth configuration $B_l$). For $2\leq l \leq M-1$ we can order the states in $\mathcal{S}_l$ as follows:
        \vspace{-0.1cm}
 		\[ \{ (R_{l-1}+1,l), \dots , (F_{l-1}+1,l), \dots , (R_l,l), \dots ,(F_l,l) \} \vspace{-0.1cm}\]
 		We define another Markov process $\mathcal{M}_l$ for $l \in \{2, \dots, M-1\}$ which corresponds to the state space $\mathcal{S}_l$. The transition structure of $\mathcal{M}_l$ resembles $\mathcal{M}$ for the states in $\mathcal{S}_l$ with the following adjustments (shown in Fig. \ref{fig:ThresholdQueueMC_2}).
 		\begin{enumerate}
 			\item A transition from $(R_{l-1}+1,l)$ to $(R_{l-1},l-1)$ in the original process $\mathcal{M}$ is replaced by a transition from $(R_{l-1}+1,l)$ to $(F_{l-1}+1,l)$.
 			\item A transition from $(F_l,l)$ to $(F_l +1, l+1)$ in the original process $\mathcal{M}$ is replaced by a transition from $(F_l,l)$ to $(R_l,l)$.
 		\end{enumerate}
        
\begin{figure}[ht]
	 			\centering
	 			\vspace{-0.3cm}
	 			\includegraphics[width=0.8\linewidth]{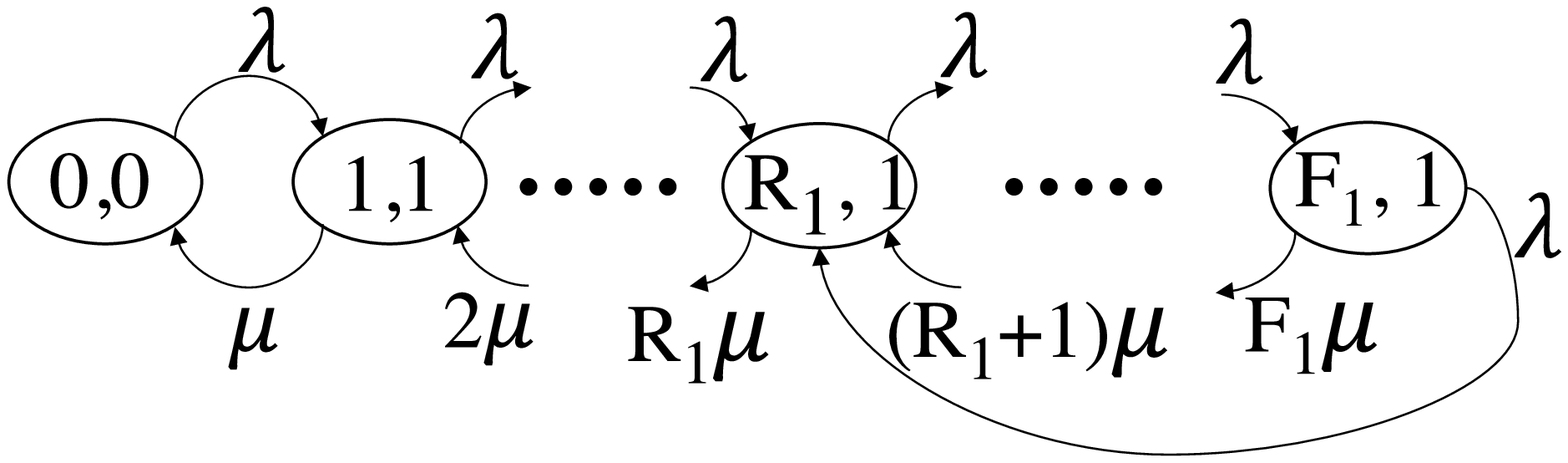}
	 			\vspace{-0.4cm}
	 			\caption{\small{State transition diagram of $\mathcal{S}_1$}}
	 			\label{fig:ThresholdQueueMC_3_p1}
	 	\end{figure}

\begin{figure}[H]
	 			\vspace{-0.3cm}
	 			\includegraphics[ width=\linewidth]{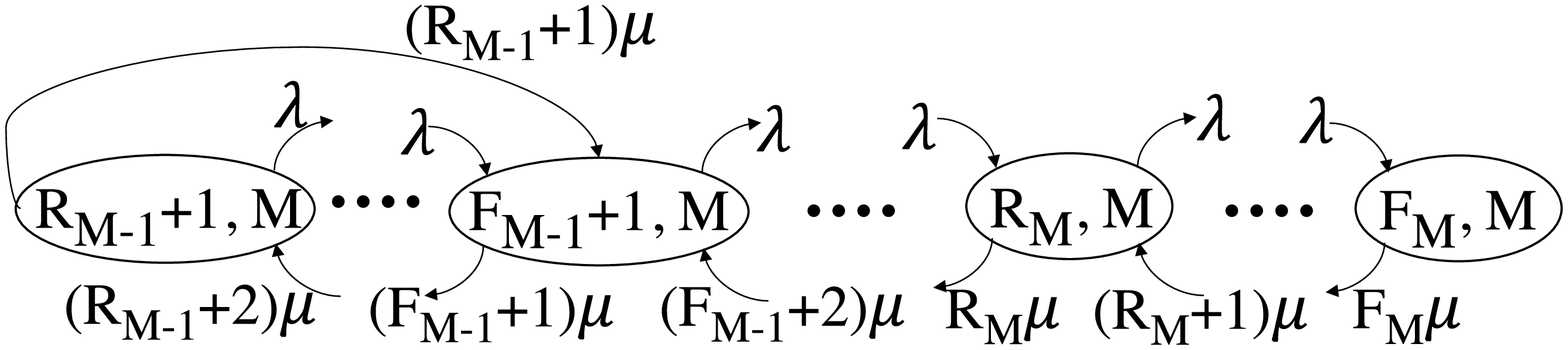}
                \vspace{-0.5cm}
	 			\caption{\small{State transition diagram of $\mathcal{S}_M$}}
	 			\label{fig:ThresholdQueueMC_3_p2}
	 			\vspace{-0.2cm}
 		\end{figure}

 			For $l=1$, adjustment (1) simply does not apply (see Fig. \ref{fig:ThresholdQueueMC_3_p1}) as there are no states to transition to from the state $(0,0)$ in the original Markov chain $\mathcal{M}$, because it is the initial state. Similarly, for $l=M$, adjustment (2) does not apply (see Fig. \ref{fig:ThresholdQueueMC_3_p2}) as $(F_M,M)$ is the terminal state of the original Markov chain $\mathcal{M}$.	
        
 		Given that we construct the partitioned Markov process $\mathcal{M}_l$ according to the procedures mentioned above, we can prove the following: \\
 		\emph{The steady-state probabilities derived from Markov process $\mathcal{M}_l$ are the conditional steady-state probabilities for the states in $\mathcal{S}_l$ of the original Markov process $\mathcal{M}$, provided that the system is in partition $S_l$.} \\
 		This statement can be proved using the argument of stochastic complementation together with the help of \emph{Theorem (1)}. The reader can refer to \cite{StochComp1} and \cite{StochComp2} for a detailed proof.
 		\newline
 		\newline
 		\emph{Analysis of $\mathcal{M}_l$}:
 		
 			Let us now derive the steady-state probability vector for the states in $\mathcal{S}_l$, namely $\pi_{l}(n)$, where $l \in \{1, \dots, M-1\}$. Based on the flow balance equation of $\mathcal{M}_l$ for $l \in \{2, \dots, M-1\}$, we can express the steady-state probabilities $\pi_l(i)$ for $(R_{l-1}+1) \leq i \leq F_l$ in terms of $\pi_l(R_{l-1}+1)$ in the following form :
 			\begin{equation}
	 			\label{eqn:Threshold_eqn_1}
	 			\begin{array}{l}
		 			\pi_{l}(i) = C_i^l\pi_{l}(R_{i-1} +1) \quad \text{for} \quad l = 2,3, \dots,M
	 			\end{array}
                \vspace{-0.3cm}
 			\end{equation} \\
 			 The unified expression of $C_i^l$ for $(R_{l-1}+1) \leq i \leq F_l$ is provided in (\ref{eqn:UnifiedThresholdRateExpression}), where $\displaystyle \rho = \lambda/\mu$. This can be obtained as follows:
             
            Considering the flow balance equation of $\mathcal{M}_l$ for $(R_{l-1}+1) \leq i \leq (F_{l-1}+1)$, we express $\pi_l(i)$ in terms of $\pi_l(R_{l-1}+1)$. Therefore, we obtain the expression of $C_i^l$ as provided in (\ref{eqn:UnifiedThresholdRateExpression_a}), for the corresponding sub-states of $\mathcal{M}_l$. 
            
            Next, we consider the flow balance equation of $\mathcal{M}_l$ for $(F_{l-1}+2) \leq i \leq R_l$ and express $\pi_l(i)$ in terms of $\pi_l(R_{l-1}+1)$. Thus, obtaining the expression of $C_i^l$ for the corresponding sub-states of $\mathcal{M}_l$ as shown in (\ref{eqn:UnifiedThresholdRateExpression_b}).
 			
 			Similarly, based on the flow balance equation of $\mathcal{M}_l$ for $(R_l+1) \leq i \leq F_l-1$, we get the expression of $\pi_l(i)$ in terms of $\pi_l(R_{l-1}+1)$ and $\pi_l(F_l)$. Therefore, we obtain the expression of $C_i^l$ for the corresponding sub-states of $\mathcal{M}_l$ as provided in (\ref{eqn:UnifiedThresholdRateExpression_c}). In (\ref{eqn:UnifiedThresholdRateExpression_c}), the expression of $C_{F_l}^l$ can be found from the from the steady-state probability of equation: $\pi_{l}(F_l) = C_{F_l}^l\pi_{l}(R_{l-1} +1)$

            \begin{figure*}[ht]
            \begin{subequations}
            \label{eqn:UnifiedThresholdRateExpression}
 				\begin{align}
 				\label{eqn:UnifiedThresholdRateExpression_a}			
 				C_i^l =				& \displaystyle \Big[ \frac{(R_{l-1}+1)!}{i!} \Big] \! \! \! \! \sum_{j=0}^{i-(R_{l-1}+1)} \! \! \! \! \! \! \rho^j \Big\{ \frac{(i \! - \! j \! - \! 1)!}{R_{l-1}!} \Big\} \qquad  \qquad \qquad  \qquad \qquad  \qquad 		 \qquad  \text{for}  \,\,\, (R_{l-1}+1) \leq i \leq (F_{l-1}+1) \\ \notag \label{eqn:UnifiedThresholdRateExpression_b}
 				\quad\,\,= 		  &\displaystyle \Big[ \frac{(R_{l-1}+1)!}{i!} \Big] \! \! \! \! \sum_{j=0}^{i-(R_{l-1}+1)} \! \! \! \! \! \! \rho^j \Big\{ \! \frac{(i \! - \! j \! - \! 1)!}{R_{l-1}!} \! \Big\} \, - \, (R_{l-1} \!+\! 1) \Big[ \frac{(F_{l-1} \!+\! 1)!}{i!} \Big] \! \! \! \! \sum_{j=0}^{i-(F_{l-1}+2)} \! \! \! \! \! \! \rho^j \Big\{ \! \frac{(i \! - \! j \! - \! 1)!}{(F_{l-1} \!+\! 1)!} \! \Big\} \\ 
 									   & \qquad  \qquad  \qquad  \qquad \qquad  \qquad  \qquad  \qquad \qquad  \quad  \qquad \qquad  \qquad \qquad  \qquad \qquad  \qquad \text{for} \quad (F_{l-1}+2) \leq i \leq R_l \\ \notag \label{eqn:UnifiedThresholdRateExpression_c}
 				\quad\,\,=		  &\displaystyle \Big[ \frac{(R_{l-1}+1)!}{i!} \Big] \! \! \! \! \sum_{j=0}^{i-(R_{l-1}+1)} \! \! \! \! \! \! \rho^j \Big\{ \! \frac{(i \! - \! j \! - \! 1)!}{R_{l-1}!} \! \Big\} \, -	 \! (R_{l-1} \!+\! 1) \Big[ \frac{(F_{l-1} \!+\! 1)!}{i!} \Big] \! \! \! \! \sum_{j=0}^{i-(F_{l-1}+2)} \! \! \! \! \! \! \! \rho^j \Big\{ \! \frac{(i \! - \! j \! - \! 1)!}{(F_{l-1} \!+\! 1)!} \! \Big\} \\
 									  &\qquad  \qquad \displaystyle - \Big[ \frac{R_l!}{i!} \Big]  \sum_{j=1}^{i-R_l} \! \!  \rho^j \Big\{ \! \frac{(i \! - \! j)!}{R_l!} \! \Big\} C_{F_l}^l \qquad  \qquad \qquad  \qquad  \qquad  \qquad  \qquad  \qquad   \text{for}  \quad (R_l+1) \leq i \leq (F_l-1)\\ \notag
 				\text{where,}  & \\ \notag \label{eqn:UnifiedThresholdRateExpression_d}
	  			C_{F_l}^l = 	  &\displaystyle \Big[ \big\{ 1 \!+\! \frac{F_l}{\rho} \big\} \!+\! \frac{R_l!}{(F_l \!-\! 1)!} \! \! \! \! \! \! \sum_{j=1}^{(F_l-1)-R_l} \! \! \! \! \! \! \rho^j \Big\{ \! \frac{((F_l \!-\! 1) \!-\! j)!}{R_l!} \! \Big\} \Big]^{-1} \Bigg[\!\!  \Big[ \frac{(R_{l-1} \!+\! 1)!}{(F_l \!-\! 1)!} \Big] \! \! \! \sum_{j=0}^{(F_l \!-\! 1)-(R_{l-1}+1)} \! \! \! \! \! \! \! \! \!\rho^j \Big\{ \! \frac{((F_l \!-\! 1) \!-\! j \!-\! 1)!}{R_{l-1}!} \! \Big\} - \\ 
	  								  &\qquad  \qquad \displaystyle (R_{l-1} \!+\! 1) \! \Big[\frac{(F_{l-1} \!+\! 1)!}{(F_l \!-\! 1)!} \Big] \! \! \! \sum_{j=0}^{(F_l \!-\! 1)-(F_{l-1}+2)} \! \! \! \! \! \! \! \! \! \rho^j \Big\{ \! \frac{((F_l \!-\! 1) \!-\! j \!-\! 1)!}{(F_{l-1} \!+\! 1)!} \! \Big\} \Bigg] \\ \notag
 				& \qquad \qquad \qquad \qquad \qquad \qquad \qquad \qquad \qquad \qquad \qquad \qquad \qquad \qquad \qquad \qquad \qquad \forall \quad l \in \{2,3, \dots, M-1\}
 				\end{align}
            \end{subequations}
            \begin{subequations} 
            \label{eqn:UnifiedThresholdRateExpression2}
 				\begin{align}
                C_i^1 =			  & \displaystyle \frac{\rho^i}{i!} \qquad  \qquad \qquad  \qquad \qquad  \qquad \qquad \qquad  \text{for}  \,\,\, 1 \leq i \leq R_1 \\
 				\quad \,\,\,	=			& \displaystyle \frac{\rho^i}{i!} - \Big[ \frac{R_1!}{i!} \Big]  \sum_{j=1}^{i-R_1} \! \!  \rho^j \Big\{ \! \frac{(i \! - \! j)!}{R_1!} \! \Big\} C_{F_1}^1 \qquad  \quad \,\,\,  \text{for}  \quad (R_1+1) \leq i \leq (F_1-1)\\
	  			 & \text{where,} \quad C_{F_1}^1 = \displaystyle \Big[ \big\{ 1 \!+\! \frac{F_1}{\rho} \big\} \!+\! \frac{R_1!}{(F_1 \!-\! 1)!} \! \! \! \! \! \! \sum_{j=1}^{(F_1-1)-R_1} \! \! \! \! \! \! \rho^j \Big\{ \! \frac{((F_1 \!-\! 1) \!-\! j)!}{R_1!} \! \Big\} \Big]^{-1} \Big[ \frac{\rho^{F_1 - 1}}{(F_1 - 1)!} \Big] 
 				\end{align}
            \end{subequations}
            \begin{subequations}
                \hspace{-2.5cm}
 				\label{eqn:UnifiedThresholdRateExpression3}			
 				\begin{align}
 				C_i^M =				& \displaystyle \Big[ \frac{(R_{M-1}+1)!}{i!} \Big] \! \! \! \! \sum_{j=0}^{i-(R_{M-1}+1)} \! \! \! \! \! \! \rho^j \Big\{ \frac{(i \! - \! j \! - \! 1)!}{R_{M-1}!} \Big\} \qquad \qquad \qquad \quad \text{for}  \,\,\, (R_{M-1}+1) \leq i \leq (F_{M-1}+1) \\ \notag
 				\qquad \! = 		  &\displaystyle \Big[ \frac{(R_{M-1}+1)!}{i!} \Big] \! \! \! \! \sum_{j=0}^{i-(R_{M-1}+1)} \! \! \! \! \! \! \rho^j \Big\{ \! \frac{(i \! - \! j \! - \! 1)!}{R_{M-1}!} \! \Big\} \, - \, (R_{M-1} \!+\! 1) \Big[ \frac{(F_{M-1} \!+\! 1)!}{i!} \Big] \! \! \! \! \sum_{j=0}^{i-(F_{M-1}+2)} \! \! \! \! \! \! \rho^j \Big\{ \! \frac{(i \! - \! j \! - \! 1)!}{(F_{M-1} \!+\! 1)!} \! \Big\} \\ 
 									   & \qquad  \qquad  \qquad \qquad \qquad \qquad \qquad  \qquad \qquad  \qquad \qquad  \qquad \text{for} \quad (F_{M-1}+2) \leq i \leq F_M
 				\end{align}
 			\end{subequations}
            \vspace{-0.3cm}
 				 \hrulefill
 			\end{figure*}
            
            Finally, considering the flow balance of $\mathcal{M}_l$ for $i=F_l$, we obtain the expression of $\pi_l(F_l)$ in terms of $\pi_l(R_{l-1}+1)$. Therefore, we obtain the expression of $C_{F_l}^l$ as shown in (\ref{eqn:UnifiedThresholdRateExpression_d}). 
            
            We follow a similar process and apply it to Markov chains $\mathcal{M}_1$ and $\mathcal{M}_M$ to derive the steady-state probabilities for $\pi_{1}$ and $\pi_{M}$, shown in (\ref{eqn:UnifiedThresholdRateExpression2}) and (\ref{eqn:UnifiedThresholdRateExpression3}), respectively. The unified expression of $C_i^l$ ($\forall\, l \in \{1,2,\dots ,M\}$) is given in equations (\ref{eqn:UnifiedThresholdRateExpression}) - (\ref{eqn:UnifiedThresholdRateExpression3}).
            
            Using (\ref{eqn:Threshold_eqn_1}), together with the newly obtained equations for $\pi_{1}$ and $\pi_{M}$, we can express $\pi_l(i)$ for $(R_{l-1}+1) \! \leq i \leq \!F_l$, $\forall \, l \in \! \{1,2,3,\dots,M\}$ ($R_0 =0$) in the following form:
 			\begin{equation}\label{eqn:UnifiedThresholdstateExpression}
            \begin{array}{l}
 				\pi_{l}(i) = C_i^l\pi_{l}(R_{i-1} +1) \quad \text{for} \quad l = 2,3,\dots,M \quad \text{and,}\\
                \pi_{1}(i) = C_i^1\pi_{1}(0)
             \end{array}
 			\end{equation}
 			where the expression for $\pi_{l}(R_{i-1} +1)$ can be obtained by summing over the state probabilities of Markov chain $\mathcal{M}_l$ and set it equal to 1.
            \begin{equation}\label{eqn:UnifiedThresholdstateExpression2}
 				\pi_{l}(R_{l-1} +1)  = \Big[ \sum_{i=0}^{F_l} C_i^l \Big]^{-1}\quad \text{for} \quad l = 1,2,\dots,M
 			\end{equation}

 			Now, let us consider the aggregated process to calculate the transition rate between the CPRI states $S_l$. Fig. \ref{fig:ThresholdQueueMC_0} shows the transition diagram of the resulting process. These are the transition rates of the RRU across the different CPRI rates. They can be computed as follows:
 			 \begin{equation} \label{eqn:ThresholdExLambdai}
	 			 \begin{array}{l}
		 			 \lambda_{i} = \lambda \pi_{i}(F_i) \qquad  \, \text{for} \quad i=1,2, \dots, M-1 \quad \text{and,} \\
		 			 \lambda_{0}=\lambda
	 			 \end{array}
 			 \end{equation}
 			 \begin{equation}\label{eqn:ThresholdExMui}
 			 	\mu_{i} = (R_{i-1}\!+\!1)\mu \pi_{i}(R_{i-1}\!+\!1)   \quad \text{for} \quad i=1,2, \dots, M
 			 \end{equation}
 			
		 \subsection{Steady-state Analysis of the Fronthaul Aggregator using Multidimensional Queuing Model}
 			 After we complete the computation of the state transition rates between different CPRI configurations, we need to analyze the steady-state probabilities at the aggregator process ($\mathcal{M}_A$). Let us consider a system with a cluster of $N$ RRUs connected to an aggregator, where each RRU adopts a set of $M$ CPRI rate configurations ($d_1, \dots, d_M$). The maximum number of RRU that can be supported with this configuration is $N^{\text{max}}_{\text{RRU}}=\left \lfloor{B_c/d_1}\right \rfloor$. The reason is that even for a single active user in the RRU, a CPRI rate of $d_1$ will be adopted. This causes the aggregated capacity to surpass the FHA link capacity, beyond a cluster size of $N^{\text{max}}_{\text{RRU}}$ RRUs. Therefore, we will always get blocking probability $P_b\approx1$.  Let $k_m$ represents the fact that RRU $k$ is using the CPRI rate configuration $d_m$. Therefore we can consider an $M$-dimensional vector $\textbf{k}=(k_1, \dots, k_m, \dots, k_M)$ to represent any steady-state of the aggregator.
 			 
\begin{figure}[h]
 			 	\includegraphics[ width =\linewidth]{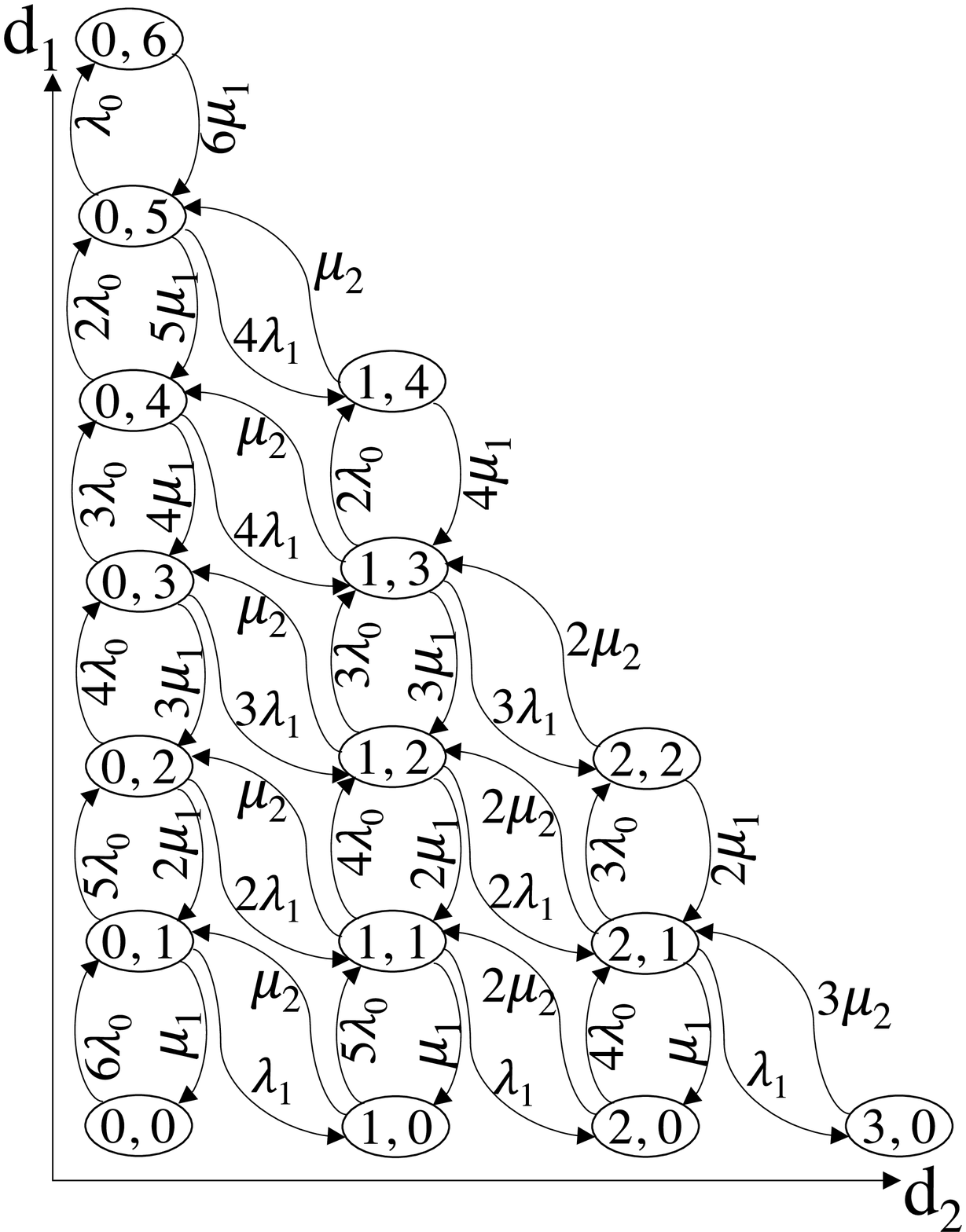}
 			 	\caption{\small{An example of the aggregator process with each RRU using two CPRI rate configurations ($d_1 \, \text{and} \, d_2$), assuming $d_2=2d_1$,  FHA link capacity = $6d_1$ and a $N$-RRU cluster connected to the aggregator ($N=6$ for this example).}}
 			 	\label{fig:AggregatorProcess}
 			 \end{figure}

 			 Depending on call arrival and departure at the RRU in the corresponding cluster, the following events can occur at the aggregator:
 			 \begin{enumerate}
 			 	\item An RRU transitions to a higher CPRI rate due to a call arrival, thus creating a transition from $d_m \to d_{m+1}$ for that RRU with a transition rate $\lambda_{m}$ ($\lambda_{m}$ is the transition rate computed in Section \ref{sec:thresholdQueue}). This creates an aggregator event for a state transition from $(k_1, \dots, k_m, k_{m+1},\dots, k_M)$ to $(k_1, \dots, k_m-1, k_{m+1}+1,\dots, k_M)$. A similar event occurs for a call departure, triggering an event at the aggregator with rate $\mu_m$.
 			 	\item A call request arrives to an RRU belonging to a cluster which had no previous user, thus creating a transition from $(d_0=0) \to d_{1}$. This is also a wake-up RRU transition with rate $\lambda$. This creates an aggregator event for a state transition from $(k_1, \dots, k_M)$ to $(k_1+1, \dots, k_M)$. A similar event occurs for a call departure, triggering an event at the aggregator with a rate $\mu_1$.
 			 \end{enumerate}
 		 	An incoming call request triggering any of these events is accepted if the new aggregated capacity, which accounts for the increase of the CPRI rate, is less than or equal to the FHA link capacity. Therefore, we get the following possible states for the aggregator:\vspace{-0.2 cm}
 		 	\begin{equation}
 		 	\vspace{-0.1 cm}
 		 	\mathbb{K} =\{\mathbf{k}\, | \,\, k_1, k_2, \dots k_M \geq 0, \sum_{i=1}^{M} k_i \leq N \,\, \text{and} \,\, \sum_{i=1}^{M} d_ik_i \leq B_c\}
 		 	\end{equation} 
 			 As there can only be one possible event at any instant of time, either due to call arrival or departure at an RRU, only a single entry of $\mathbf{k}$ can change at any epoch. Therefore, we can consider the aggregator as a multi-dimensional birth and death process. In the following text, we provide the mathematical description of the transition from the $i^{\text{th}}$ to the $j^{\text{th}}$ state of the aggregator process $\mathcal{M}_A$.
 			 
 			 \begin{equation}
 			 \label{eqn:aggregatorTransition}
	 			 \begin{array}{l}
		 			 \!\!\!\!\! \displaystyle Q_{\mathbf{k}^{(i)}\mathbf{k}^{(j)}} \vspace{0.3cm} \\
		 			 \displaystyle \quad = \mathbf{k}_m^{(i)}\lambda_m  \qquad \qquad \,\,\,\, \text{if} \quad \mathbf{k}^{(j)} -\mathbf{k}^{(i)} = e_m^{(1)}, \,\, \mathbf{k}_m^{(i)} \neq 0 \vspace{0.2cm}\\
		 			 \displaystyle \quad = \!\! \Big(\! N \!-\!\! \sum_{m=1}^{M}\!\!\mathbf{k}_m^{(i)} \!\Big)\!\lambda \,\,\,\, \quad \text{if} \,\,\, \mathbf{k}^{(j)} \!\! -\mathbf{k}^{(i)} = e_1^{(2)},  \sum_{m=1}^{M}\!\mathbf{k}_m \!<\! N , \vspace{0.1cm}\\
		 			 \displaystyle \qquad  \qquad   \qquad  \qquad   \qquad  \qquad   \qquad  \qquad  \text{and } \, N\leq N^{\text{max}}_{\text{RRU}}\vspace{0.2cm}\\
		 			 \displaystyle \quad = \mathbf{k}_m^{(i)}\mu_m  \qquad \qquad \,\,\, \text{if} \quad \mathbf{k}^{(j)} -\mathbf{k}^{(i)} = -e_m^{(1)}, \,\, \mathbf{k}_m^{(i)} \neq 0\vspace{0.2cm}\\
		 			 \displaystyle \quad = \mathbf{k}_1^{(i)}\mu_1  \qquad \qquad \,\,\,\,\, \text{if} \quad \mathbf{k}^{(j)} -\mathbf{k}^{(i)} = -e_1^{(2)}, \,\, \mathbf{k}_1^{(i)} \neq 0\vspace{0.2cm}\\ 
		 			 \displaystyle \quad = 0 \qquad \qquad \qquad \,\,\,\,\, \text{Otherwise}
	 			 \end{array}
 			 \end{equation}
 			 where states $\mathbf{k}^{(i)} ,\mathbf{k}^{(j)} \in \mathbb{K}$, $\mathbf{k}^{(i)}_m$ is the $m^{\text{th}}$ entry of $\mathbf{k}^{(i)}$. 
 			 \begin{equation*}
	 			 \begin{array}{lll}
		 			 &\qquad \qquad \, m^{\text{th}}&\!\!\!\!\!(\!m\!\!+\!\!1\!)^{\text{th}} \vspace{-0.17cm}\\
		 			{e_m^{(1)}}^T =& \{0,0, \dots, -1 \,\,,& 1, \dots , 0, 0 \} \qquad \text{and} \\
		 			{e_1^{(2)}}^T =&  \{1,0, \dots, 0, 0 \}& \\
	 			 \end{array}
 			\end{equation*}
 			 An example of the aggregator process is shown in Fig. \ref{fig:AggregatorProcess}.
 			 \newline
 			 \newline
 			 \emph{Analysis of $\mathcal{M}_A$}:
 			 
 			 It can be shown that $\mathbf{k}$ satisfies the reversibility property by following the method described in \cite{kaufman_blocking_1981}. We use this property to derive the steady-state probabilities at the aggregator and the obtained expression is given in (\ref{eqn:steadystLocBalEqn_9}).
 			\begin{equation}\label{eqn:steadystLocBalEqn_9}
 			\begin{array}{l}
 			\displaystyle \!\!\mathop{P}(k_1, k_2, \dots,  k_M) =   \vspace{0.3cm} \\
 			\displaystyle \quad \mathop{P}(0, 0, \dots, 0)\Bigg[\!\!\binom{\!N_{\text{RRU}}\!}{K_s} \frac{K_s!}{\prod_{i=1}^{M}k_i!} \!  \prod_{i=1}^{M} \!\! \bigg(\!\!  \frac{\lambda_{i-1}}{\mu_i}  \!\! \bigg)^{\!\!\sum_{j=i}^{M}k_j} \Bigg] \!\!\!\!
 			\end{array}
 			\end{equation}
 			In (\ref{eqn:steadystLocBalEqn_9}), $N_{\text{RRU}}=N$, for the cluster size $N \leq N^{\text{max}}_{\text{RRU}}$ and $N_{\text{RRU}}=N^{\text{max}}_{\text{RRU}}$, for the cluster size $N > N^{\text{max}}_{\text{RRU}}$.
 			The reader is invited to see APPENDIX A for the derivation of (\ref{eqn:steadystLocBalEqn_9}).\\
 			 We can derive the expression for $P(0, 0, \dots, 0)$ as
 			 \begin{equation}\label{eqn:steadystLocBalEqn_10}
 			 \begin{array}{l}
 			 \displaystyle \mathop{P}(0, 0, \dots, 0) =   \vspace{0.3cm} \\
 			 \displaystyle \quad \Bigg[ \sum_{\mathbf{k} \in \mathbb{K}}^{} {\bigg[\!\!\binom{N_{\text{RRU}}}{K_s} \frac{K_s!}{\prod_{i=1}^{M}k_i!}   \prod_{i=1}^{M} \! \bigg(\!  \frac{\lambda_{i-1}}{\mu_i}  \! \bigg)^{\!\sum_{j=i}^{M}k_j} \bigg]} \!\Bigg]^{\!-1}\!\!\! 
 			 \end{array}
 			 \end{equation}
 			 by using the fact that
 			 \[\sum_{\mathbf{k} \in \mathbb{K}}\mathop{P}(k_1, k_2, \dots, k_m, \dots, k_M) =1\]
 			 
 			 Therefore, the steady state probabilities of the aggregator can be obtained from (\ref{eqn:steadystLocBalEqn_9}). In this expression, the transition rates between different CPRI configurations $\lambda_{i}$ and $\mu_{i}$ can be obtained from (\ref{eqn:UnifiedThresholdstateExpression})  - (\ref{eqn:ThresholdExMui}).
 			 
		 \subsection{Blocking Probabilities at the Fronthaul Aggregator}
		 	With the expression of steady-state probabilities derived in the previous section, we can evaluate the blocking probability at the aggregator. In order to accomplish this, we can decompose the calculation of blocking probability into $M$-parts with $M$ being the number of CPRI rates (or bandwidth configurations) available at the RRUs belonging to a cluster. Let $P_B^{\lambda_{m}}$ be the probability that the aggregator goes into a blocking state due to a transition of CPRI rate from $d_m$ to $d_{m+1}$. We take $d_0 =0$ and $\lambda_{0} = \lambda$ as we have described previously. Then $P_B^{\lambda_{m}}$ can be determined as follows:
		 	\begin{equation}
		 		\begin{array}{l}
		 			\displaystyle\!\!	P_B^{\lambda_{m}} = \frac{ \displaystyle k_m \lambda_m \!\! \sum_{\mathbf{k} \in \mathbb{K}^{\lambda_m}} P(\mathbf{k})  } {\displaystyle \sum_{i=0}^{M-1} \lambda_{m}} \qquad \text{for} \,\, i =1,2, \dots, M\!-\!1 \!\!\!\!\! \vspace{0.3 cm}\\
		 			\!\!\!\!\!\!\text{and} \\
		 			\displaystyle\!\!	P_B^{\lambda_{0}} = \frac{ \displaystyle \Big( N_{\text{RRU}}-\sum_{i=1}^{M}k_i \Big) \lambda \!\! \sum_{\mathbf{k} \in \mathbb{K}^{\lambda_0}} P(\mathbf{k})  } {\displaystyle \sum_{i=0}^{M-1} \lambda_{m}}
		 		\end{array}
		 	\end{equation}
		 	where,  \\ $\displaystyle \mathbb{K}^{\lambda_{m}} \!= \{ \mathbf{k} \,| \quad \mathbf{k} \in \mathbb{K}, k_m \neq 0, \text{and} \sum_{i=1}^{M}\!(k_i \!+\! e_{m\_i}^{(1)})d_i \!>\! B_c\}$,  $\vspace{0.4cm}\displaystyle \mathbb{K}^{\lambda_{0}} \!= \{ \mathbf{k} \,| \quad \mathbf{k} \in \mathbb{K}, \sum_{i=1}^{M}k_i \!<\! N, \text{and} \sum_{i=1}^{M}\!(k_i \!+\! e_{1\_i}^{(2)})d_i\}$,  
            
            $e_{m\_i}^{(1)}$ is the $i^{\text{th}}$ entry of $e_{m}^{(1)}$ and $e_{1\_i}^{(2)}$ is the $i^{\text{th}}$ entry of $e_{1}^{(2)}$.
		 	
		 	\vspace{0.2cm}
		 	Finally, the overall blocking probability at the aggregator can be found as
		 	\begin{equation}
			 	P_B = \sum_{m=0}^{M-1}P_B^{\lambda_m} 
		 	\end{equation} 
		 	
\section{Simulation Model} \label{sec:SimulationModel}
\begin{figure}[h]
		\includegraphics[clip, trim={0, 2.5in, 0, 0}, width=1.05\linewidth]{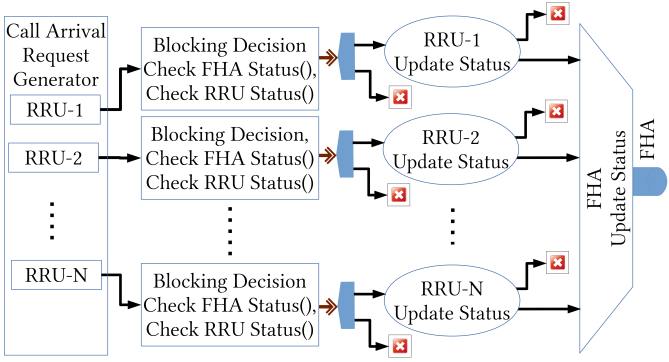}
		\caption{Model of the variable rate fronthaul simulator implemented in MATLAB}
		\label{fig:SimulationBD}
	\end{figure}

	In addition to providing an analytical expression for the blocking probability at the fronthaul aggregator, we have also simulated the above model in Matlab using the discrete-event-blocks called `Sim Events'. The model is shown in Fig. \ref{fig:SimulationBD}. For each RRU belonging to an RRU-cluster, the call arrival and service is simulated as an $M/M/S(0)$ loss system queuing process. In this work, we have taken the number of servers $S$ as $50$, as shown in table \ref{tab:CPRIdataRates}, which corresponds to the number of RGs available at the $20$ MHz LTE bandwidth configuration. We use an event-based function call module to translate the current state of the RRU to its corresponding fronthaul rate. In this simulation, an incoming call is dropped whenever the sum of all fronthaul rates at the aggregator is higher than the aggregator capacity. In the case of an XG-PON system, this is equivalent to 10 Gbps. The events in the aggregator are triggered by the events in any RRU. Therefore, an incoming call at any RRU can be blocked for two reasons:
	\begin{enumerate}
		\item If there is no idle server available at the call arrival instance to serve the incoming request, i.e., all 50 servers are busy. This means no RG is available to serve the call, which is thus blocked and dropped by the RRU.
		\item If due to a call arrival, a higher bandwidth configuration needs to be adopted but there is not enough capacity available in the FHA link. Therefore the call request is blocked due to unavailable bandwidth at the FHA link.
	\end{enumerate}
	%
		Tables \ref{tab:FronthaulConfig} and \ref{tab:SimulationPar} provide the parameters used in our simulations.  If the holding time of a service session (or call duration of an accepted call) is $\tau$ time units, the service rate ($\mu$) corresponding to each service session of a particular RRU is: $\frac{1}{\tau}$ (calls/time units). If the maximum number of service sessions per RRU is `$\mathcal{K}$' then the condition for the call arrival rate ($\lambda$) in each RRU so that the system maintains a steady state blocking probability is  $\frac{\lambda}{\mathcal{K} \mu} < 1$ \cite{akimaru_teletraffic_2012}. 
	The traffic load (in Erlang) for each RRU can be represented as $\rho = \lambda/\mu$. Hence, we can say the traffic load can essentially go up to $\mathcal{K}$ Erlang. Therefore, we denote $a = \frac{\lambda}{\mathcal{K}\mu}$ in order to obtain a normalized traffic load across RRUs. We use different values of `$a$' to control the traffic load per RRU.
	\begin{table*}[t]
		\vspace{-0.1in}
		\caption{Fronthaul rates used for different VBF configuration}
		\label{tab:FronthaulConfig}
		\centering
		\setlength\tabcolsep{1.5pt}
		\begin{tabular}{|>{\centering\arraybackslash}m{0.127\textwidth}|>{\centering\arraybackslash}m{0.21\textwidth}|>{\centering\arraybackslash}m{0.23\textwidth}|>{\centering\arraybackslash}m{0.3\textwidth}|}
			\hline
			$N_d = 1$ & $N_d = 2$ &$N_d = 3$ &$N_d = 4$ \\ \hline
			1228.8 Mbps&\{614.4, 1228.8\} Mbps&\{307.2, 614.4, 1228.8\} Mbps&\{153.6, 307.2, 614.4, 1228.8\} Mbps\\ \hline
		\end{tabular}
		\vspace{-0.1in}
	\end{table*}
	
	\begin{table}[h]
		\vspace{-0.1in}
		\caption{Simulation parameters}
		\label{tab:SimulationPar}
		\setlength\tabcolsep{1.5pt}
		\begin{tabular}{|p{0.75\linewidth}|>{\centering\arraybackslash}p{0.2\linewidth}|}
			\hline
			Service Rate (per service session)& 0.5\\ \hline
			Max no. of service sessions per RRU& 50\\ \hline
			No. of available fronthaul rates ($N_d$) & 1,2,3,4\\ \hline
			Capacity of FHA link                                 & 10 Gbps\\ \hline
			Max no. of RRU per Aggregator         &20\\ \hline
			Number of servers in Aggregator            &20\\ \hline
		\end{tabular}
		\vspace{-0.1in}
	\end{table}
	\vspace{-0.08in}
	
\section{Results} \label{sec:Results}
	Here we report and compare the results of both our theoretical and simulation analysis of our variable rate fronthaul system, showing the blocking probability for a number of different configurations. We run the event-driven simulations for approximately $10^{7}$ events for each specific system configuration (i.e., for a combination of $M$, $N$ and $a$) and capture the blocking probability. 
	
\begin{figure}[h]
		\includegraphics[clip, trim={0.0in, 0, 0.2in, 0.4in }, width=\linewidth]{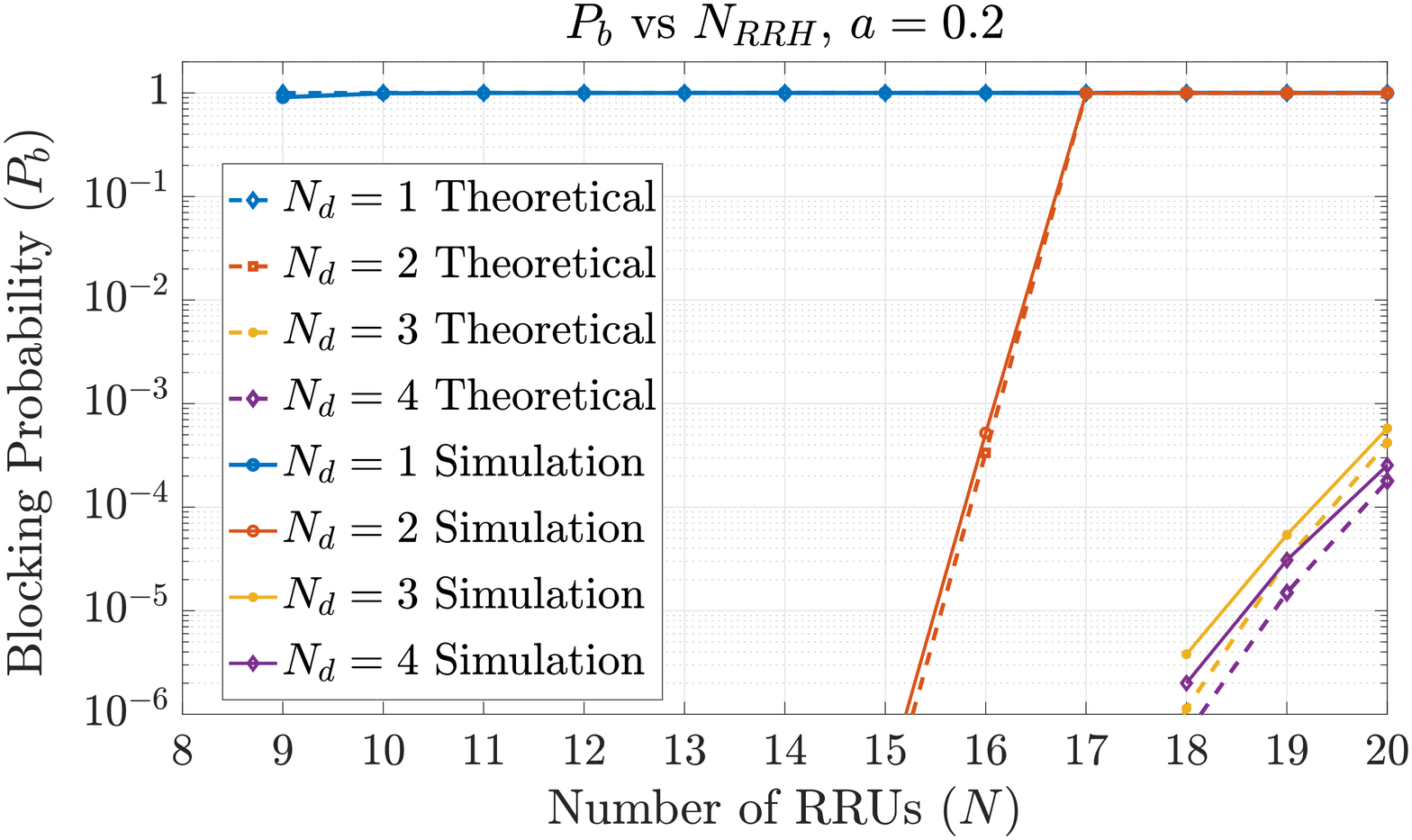}
		\vspace{-0.2in}
		\caption{\small{Blocking probability ($P_b$) \emph{vs.} number of RRUs for $a=0.2$}}
		\label{fig:Pb_Vs_numRRH_a_0.2}
	\end{figure}

\begin{figure}[h]
		\includegraphics[clip, trim={0.1in, 0, 1.5in, 0.5in },width=\linewidth]{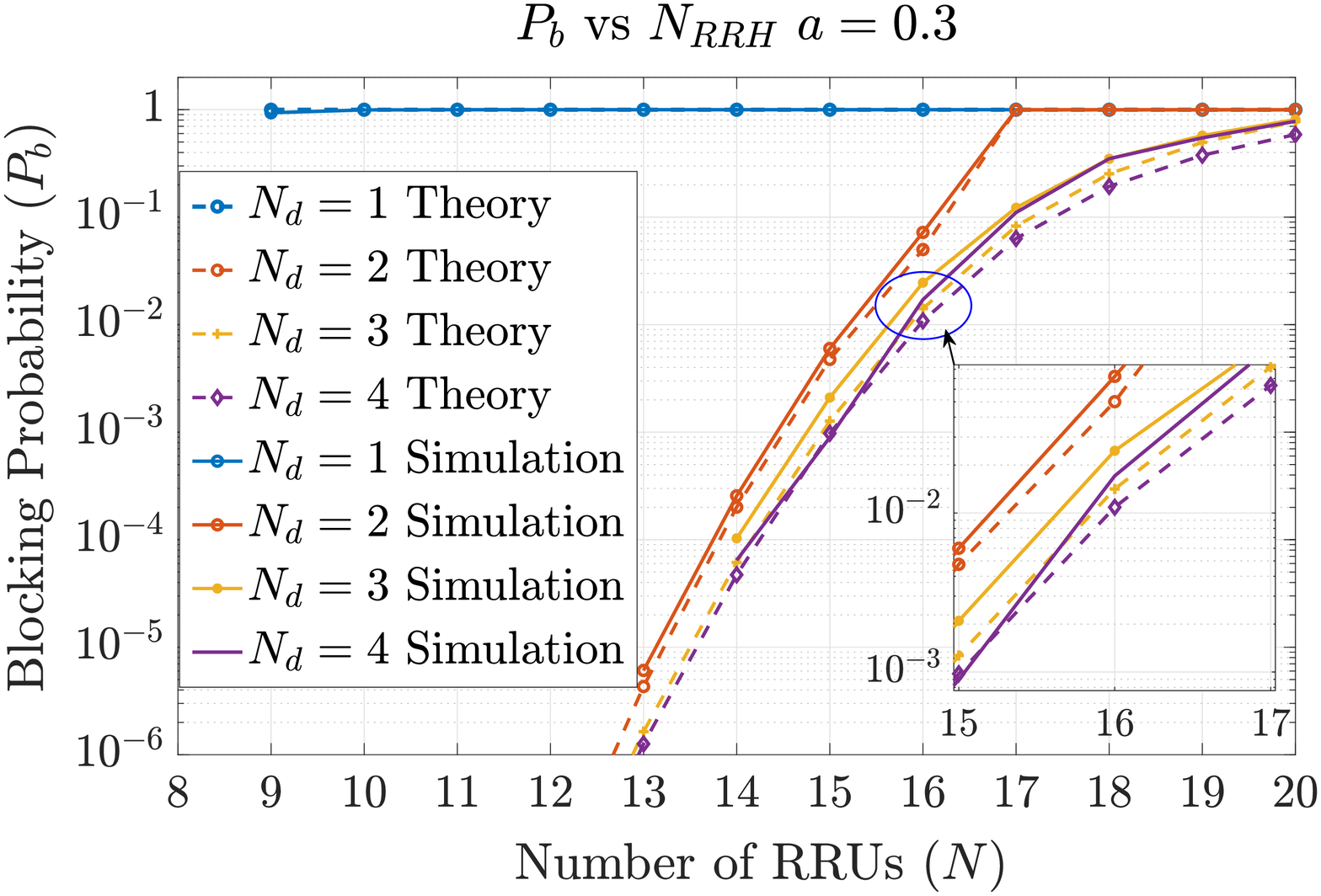}
		\vspace{-0.2in}
		\caption{\small{Blocking probability ($P_b$) \emph{vs.} number of RRUs for $a=0.3$}}
		\label{fig:Pb_Vs_numRRH_a_0.3}
		\vspace{-0.15in}
	\end{figure}

\begin{figure}[h]
		\includegraphics[clip, trim={0.05in, 0, 0.7in, 0.5in },width=\linewidth]{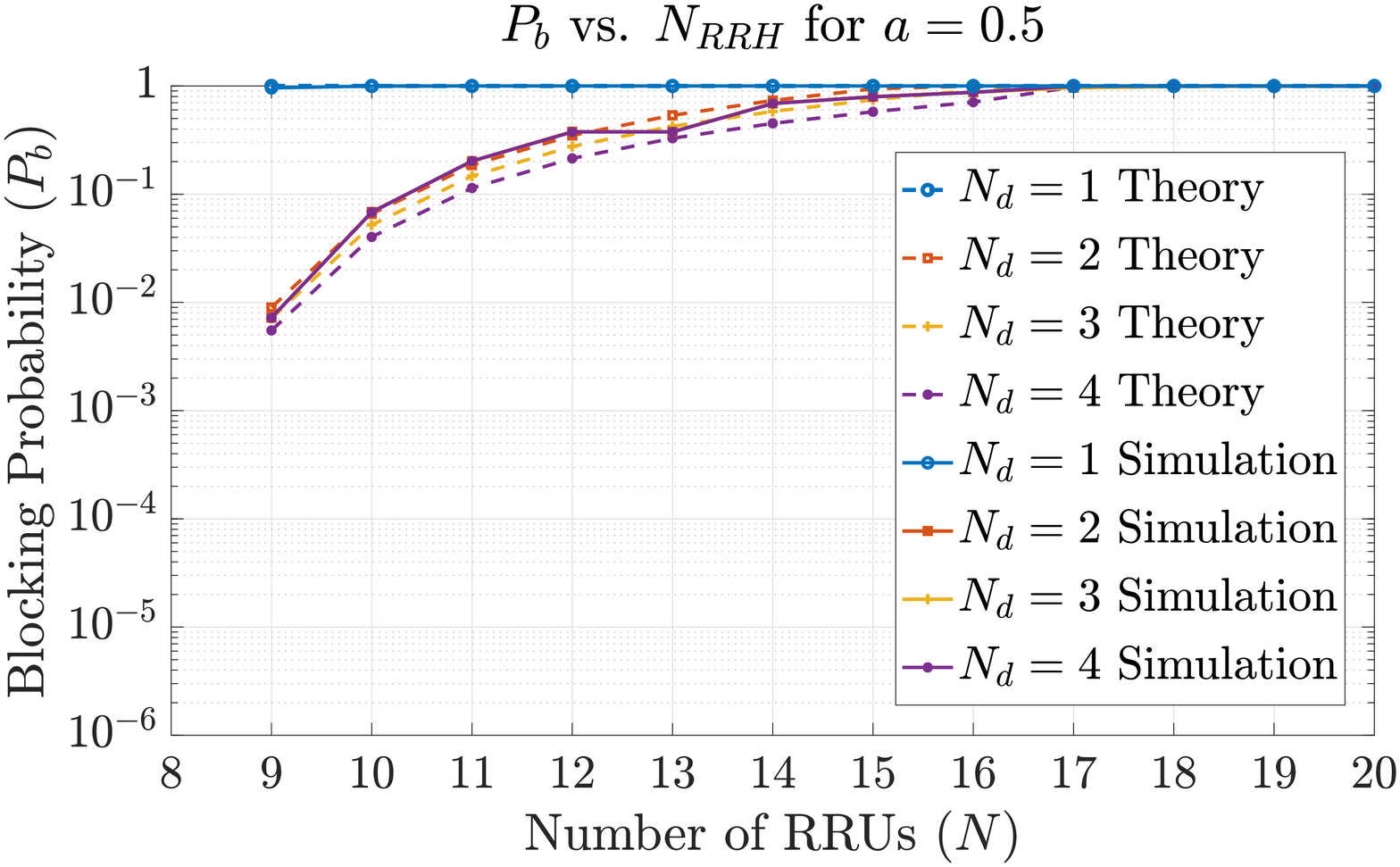}
		\vspace{-0.2in}
		\caption{\small{Blocking probability ($P_b$) \emph{vs.} number of RRUs for $a=0.5$}}
		\label{fig:Pb_Vs_numRRH_a_0.5}
	\end{figure}

	Fig. \ref{fig:Pb_Vs_numRRH_a_0.2} to \ref{fig:Pb_Vs_numRRH_a_0.5} illustrate the blocking probability ($P_b$) w.r.t the RRU cluster size for traffic intensity (`$a$') of 0.2, 0.3, and 0.5, respectively. In these figures, $N_d$ represents the number of different data rates that can be used for the fronthaul transport (i.e., the parameter $M$, as defined in the theoretical analysis section). Therefore, $N_d=1$ corresponds to the traditional fronthaul scheme. These figures also demonstrate the improvement of blocking probability when our proposed variable rate fronthaul scheme is used over traditional fronthaul (i.e., for $N_d=1$). In these results, we have taken the forward thresholds ($F_l$) as the maximum number of supported users corresponding to the LTE bandwidth / CPRI rate configuration (as listed in Table \ref{tab:CPRIdataRates}) and the reverse thresholds ($R_l$) as one step less than the $F_l$. Therefore, the difference between the forward and reverse threshold is equal to 1. 
    
   Firstly, we can see a close match between the analytical and simulated results, which corroborates the validity of our analysis. We do notice a slight discrepancy at the low end of the blocking probability ($\approx 10^{-4} - 10^{-5}$), which we believe is a statistical deviation due to the low number of instances where blocking occurred. 
    In Fig. \ref{fig:Pb_Vs_numRRH_a_0.2}, it can be observed that for the traditional fronthaul scheme the blocking probability is zero up to a cluster of size 8. After that, it sharply increases to 1. This is because, in traditional fronthaul ($N_d=1$), each of the RRU adopts a static rate of 1228.8 Mbps even for a single active user in the system, causing the aggregated rate to surpass FHA link capacity of 10 Gbps.
	In addition, for a cluster size of 9, we notice that $P_b$ is slightly less than 1 ($\approx$ 0.9). The reason for this is that at moderate traffic intensity ($a=0.2$), it is relatively likely that there are no users in at least one RRU at any time instant. In that case, we have assumed that the RRU does not send any data over the fronthaul link. Thus, only 8 RRUs would be active, which is a non-blocking situation. 
    
	
    
     Secondly, it is clear from the figures that our proposed variable rate fronthaul scheme achieves noticeable lower blocking probability compared to the static CPRI case. Indeed for values of $a = 0.2$, we can see that a blocking probability below $10^{-4}$ can be achieved even when 15 cells are aggregated (i.e., almost the double of using traditional fronthaul). The number of cells can be increased to 18 if four different rates are employed. However, it should also be noticed from Fig. \ref{fig:Pb_Vs_numRRH_a_0.3} that we obtain almost the same performance for $N_d=3$ and $N_d=4$ for $a=0.3$ (the difference is highlighted in the small square). This is due to the fact that in our system the different CPRI rates are considered in descending order and, for example, the $N_d=4$ configuration only adds the low 153.6 Mbps data rate with respect to $N_d=3$, which has a small impact on the aggregated rate. Furthermore, the maximum number of end users supported at the 153.6 Mbps rate is only 6, meaning that if more than six calls arrive at the RRU, this will move to a higher data rate. Thus, the probability for the RRU to remain in this lowest rate is significantly small. The difference can only be noticed when either the call arrival rate is very low or the size of the RRU cluster is large. As we increase the traffic intensity, for $a=0.5$ (and for rates above $a=0.5$, which are not shown here), there is very little difference in performance for $N_d=2$, $N_d=3$ and $N_d=4$. This is illustrated in Fig. \ref{fig:Pb_Vs_numRRH_a_0.5} and 
	  happens because at this high call arrival rate, higher rate configurations (614 Mbps and 1228.8 Mbps) are adopted most of the time. 
      
\vfill{}

\begin{figure}[h]
		\includegraphics[clip, trim={0.1in, 0, 0.9in, 0.5in }, width=\linewidth]{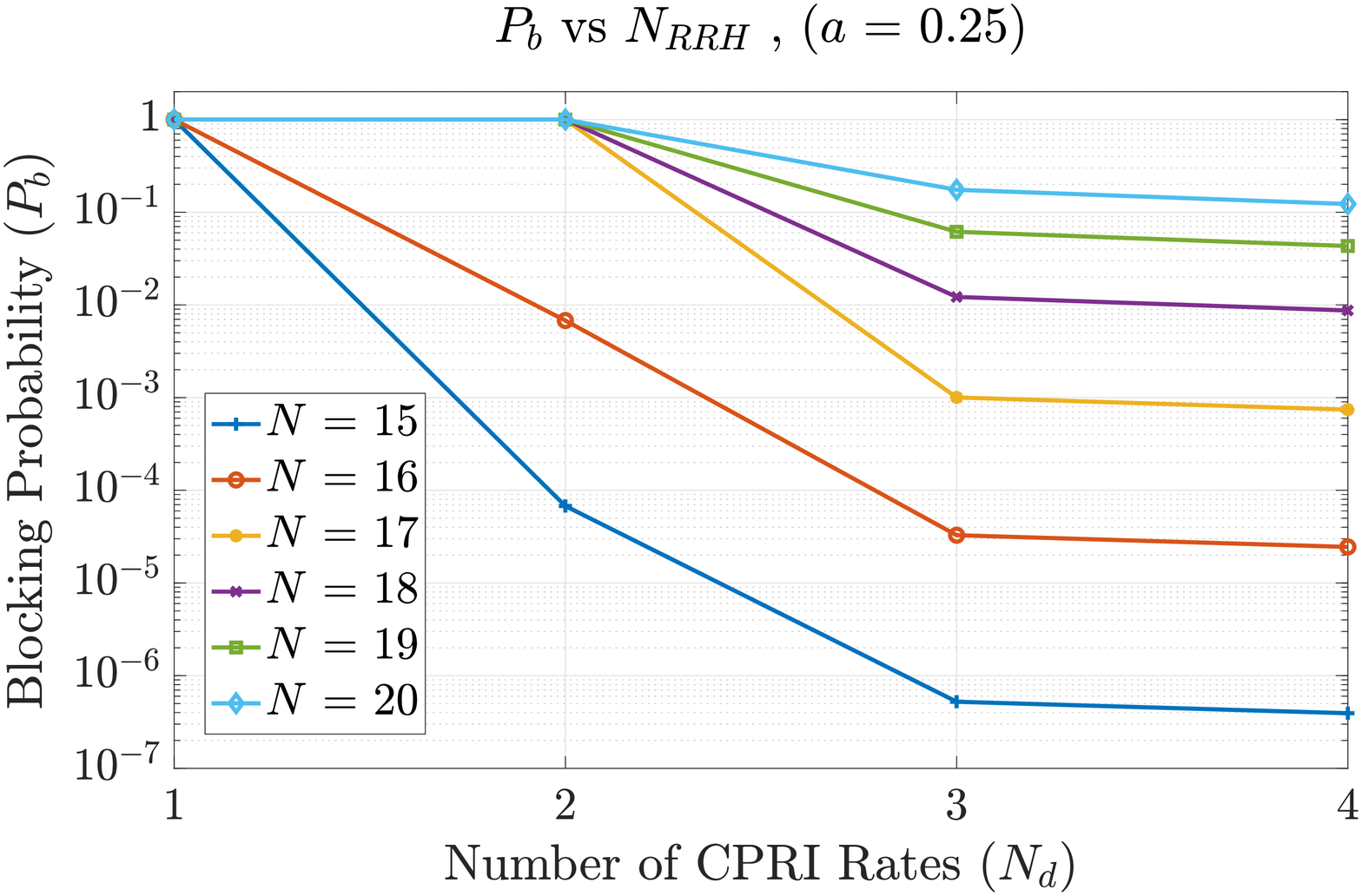}
		\vspace{-0.2in}
		\caption{\small{Blocking probability ($P_b$) \emph{vs.} number of data rates for $a=0.25$}}
		\label{fig:Pb_Vs_Nd_a_0_25}
	\end{figure}

	 In Fig. \ref{fig:Pb_Vs_Nd_a_0_25}, we use our results to  determine the maximum cluster size for which a given blocking probability can be achieved, for $a=0.25$. For example, a blocking probability of $10^{-3}$ can be attained with a cluster size of up to 17 RRUs, if we use a VRF configuration with three CPRI rates. However, if we only use two CPRI rates, then the maximum cluster size is reduced to 15.

\begin{figure}[h]
		\includegraphics[clip, trim={0.0in, 0, 0.5in, 0.6in }, width=\linewidth]{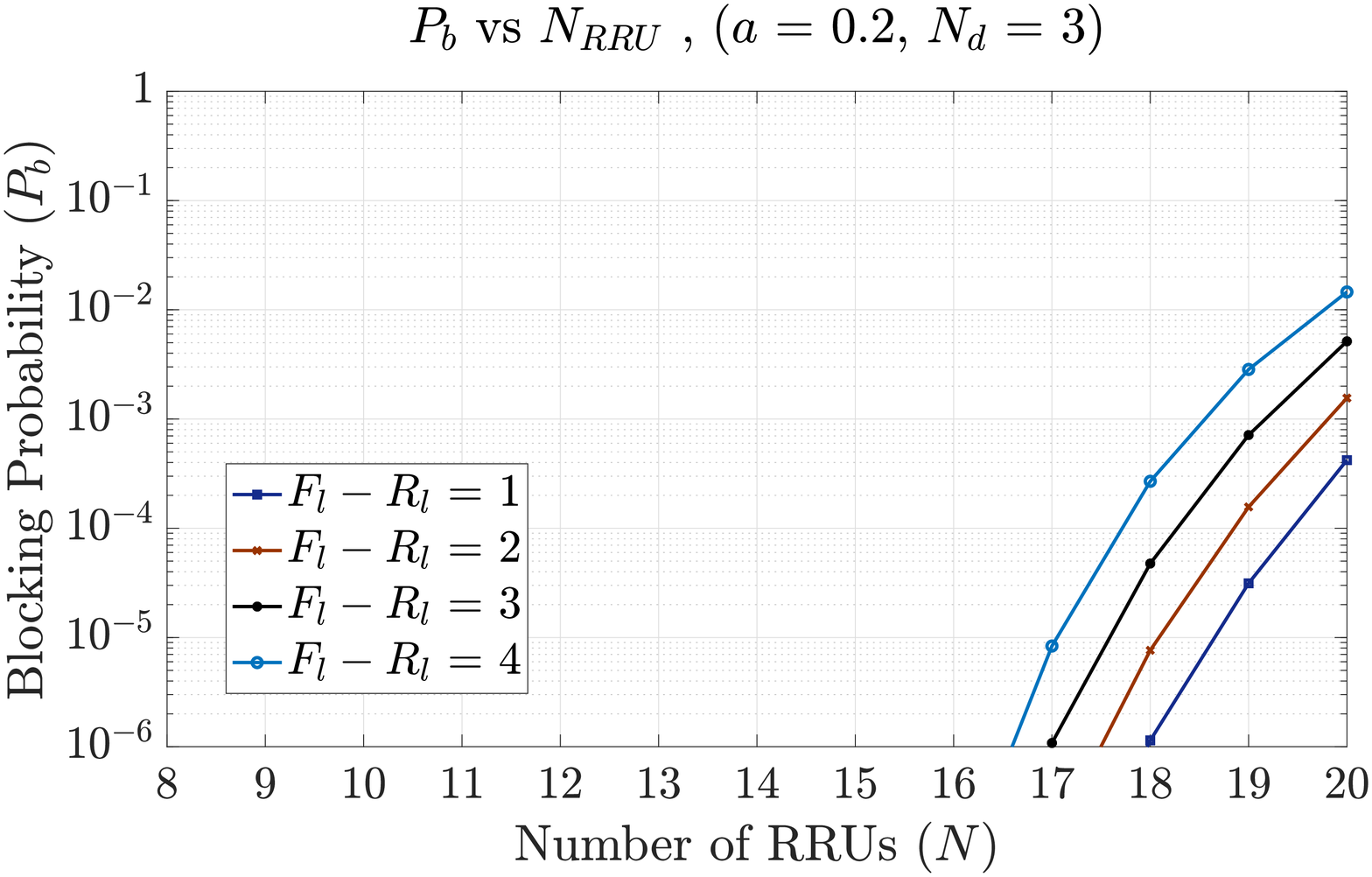}
		\vspace{-0.3in}
		\caption{\small{Blocking probability ($P_b$) \emph{vs.} number of RRUs for various differences between forward and reverse thresholds ($F_l - R_l$ =1,2,3,4), for $a=0.2$ and $N_d=3$}}
		\label{fig:Pb_Vs_numRRH_ThresholdDiffComp}
	\end{figure}

	The results shown up to this point have considered a difference between the forward ($F_l$) and reverse threshold ($R_l$) of 1. Fig. \ref{fig:Pb_Vs_numRRH_ThresholdDiffComp} shows how the blocking probability increases when we increase the difference between these thresholds. This result considers normalized traffic load as $a=0.2$, where each RRU employs three different CPRI rates ($N_d=3$). Indeed, as we increase such difference ($F_l - R_l$), a given CPRI state is retained for a longer period, till a higher number of call departs from the system (i.e., with respect to the case where $F_l - R_l =1$). This result is important when considering the finite amount of time required for a mobile system to transition between different CPRI rates. 
	
	 The difference between the forward and reverse threshold helps to prevent hysteresis i.e., to prevent looping between two CPRI configurations. Furthermore, it also takes care of the latency encountered by the SDN controller to configure between different CPRI rates. For example, if the latency for reconfiguration between different CPRI rates is somewhere between 500ms (this is a value we experimented in our dynamic-bandwidth SDR testbed implementation \cite{VBF_OFC_paper}) to 5seconds (this is the average value taken to completely reboot the SDR BBU) and if we consider normalized traffic load (for example $a=0.2$) per RRU then we can write,\\
	\noindent $a=\lambda/\mathcal{K}\mu = 0.2$\\
	$\implies \lambda=0.2\times\mathcal{K}\times\mu=10\mu$ ($\mathcal{K}=50 \rightarrow$ maximum number of users supported)
	
	 Now, if we consider a moderate traffic scenario, with $\lambda$ = 10 calls/min (e.g., in terms of number of users joining the cell) and $\mu$ = 1 calls/min, this means that one call, two calls and three calls are expected to arrive, within the reconfiguration window (0.5 seconds), respectively, with probabilities $(p)=0.0767, 0.0032$ and $8.8739\times10^{-5}$. Therefore, if we choose $F_l-R_l < 2$ (say for example $F_l-R_l = 1$), then while the SDN controller is triggering a transition to a lower CPRI rate, it is highly probable that one or more calls might arrive. This immediately triggers a transition to a higher CPRI rate which makes the system unstable. Therefore  $F_l-R_l \geq 2$ is a good choice to efficiently address the hysteresis (keeps it within 1 percent) and SDN controller configuration timings. However, if instead of implementing a dynamic reconfiguration mechanism, the BBU needs a reboot, with a reconfiguration timing window of 5 seconds, then using the same argument we see that for $a=0.2, \lambda=10 \text{ and } \mu=1$, the probability of arrival of two, three and four call requests is $0.1509, 0.0419$ and $0.0087$ respectively. Therefore a choice of $F_l-R_l \geq 3$ efficiently keeps the hysteresis within 1 percent for this case.  
	
	It should be noted that the choice of the parameter $F_l-R_l$ only depends on the arrival and departure rates considered in the system. Furthermore, given an average traffic load with moderate intensity, if we let the reconfiguration window to be higher, then the system will be more prone to stay in the higher CPRI rate and the lower rates will be used fewer times, thereby reducing the advantage of using more CPRI configurations ($N_d$) in VRF. However, as the traffic load increases, the lower rates would be used less frequently in any case (as shown in Fig. \ref{fig:Pb_Vs_numRRH_a_0.2}-\ref{fig:Pb_Vs_numRRH_a_0.5}), therefore reducing the effect of the reconfiguration window on the system performance. 
	
	Finally, while in the future we expect C-RAN software-based BBUs to all have similar reconfiguration times to the values above, today some hardware-based BBU might have notably longer reconfiguration times, in which case the rate should be modified less frequently, based on predicted average traffic.
    
\begin{figure}[h]
    	\includegraphics[clip, trim={0.0in, 0, 0.5in, 0in }, width=\linewidth]{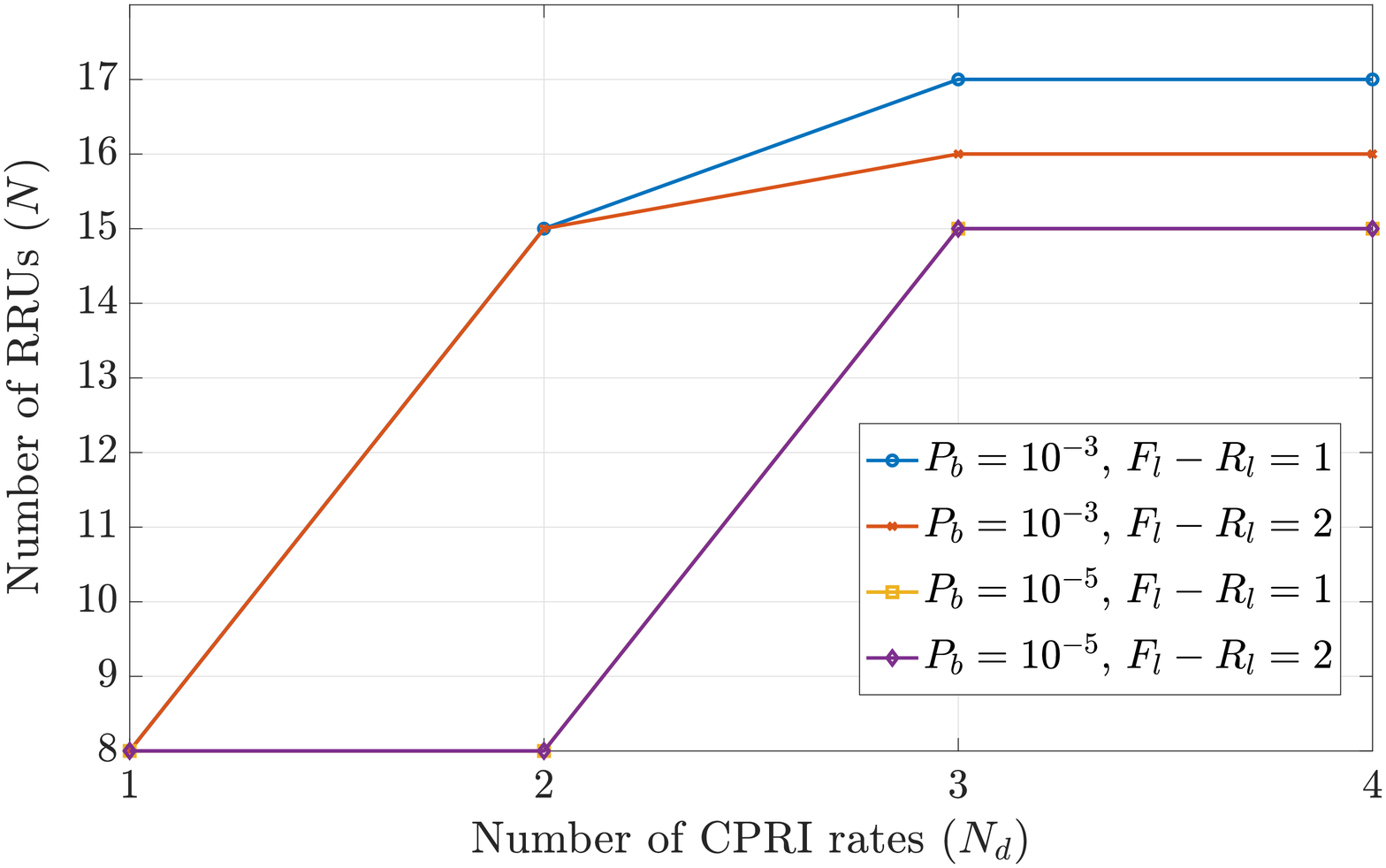}
    	\vspace{-0.3in}
    	\caption{Maximum number of RRUs that can be aggregated for different VRF configuration under certain Grade of Service ($P_b =10^{-3}, 10^{-5}$) requirement, normalized traffic load ($a=0.25$), and different choice of forward and reverse threshold difference ($F_l-R_l =1,2$).}
    	\label{fig:Comp_Nrrh_Nd_a_0_25}
    \end{figure}

    Fig. \ref{fig:Comp_Nrrh_Nd_a_0_25} illustrates a typical deployment scenario that could be obtained summarizing the results from Fig. \ref{fig:Pb_Vs_numRRH_a_0.2} to Fig. \ref{fig:Pb_Vs_numRRH_ThresholdDiffComp}. This result provides the number of RRUs that can be aggregated for different VRF configurations ($N_d =1,2,3,4$) under the requirement of a certain Grade of Service (GoS) (or blocking probability ($P_b = 10^{-3},  10^{-5}$)) while considering a given normalized traffic load ($a=0.25$) and a choice of the forward and reverse threshold difference ($F_l-R_l =1,2$). For $N_d=1$, the system operates over traditional CPRI, so different values of $F_l-R_l$ don't make any difference. For $N_d=2$ however, there is a difference between $F_l-R_l=1$ and $F_l-R_l=2$ for $P_b=10^{-5}$.

    In our study, the Poisson arrival is linked to the increase of PRBs as more users join the network, which can be described as a Poisson process. However, it is true that additional PRBs could be allocated to the same user, when its requested capacity increases thus making the traffic non-Poisson in nature. The theoretical analysis for these additional cases could not be carried out because for non-Poisson distribution the system cannot be decoupled via partitions. In this paper, we carried out simulation using two non-Poisson traffic namely, Weibull arrival process with shape factor ($k$)$=0.9 \text{ and } 1.5$ respectively. A summary of this distribution is provided in APPENDIX B
    
    \begin{figure}[h]
		\includegraphics[clip, trim={0.0in, 0.0in, 0.5in, 0.7in }, width=\linewidth]{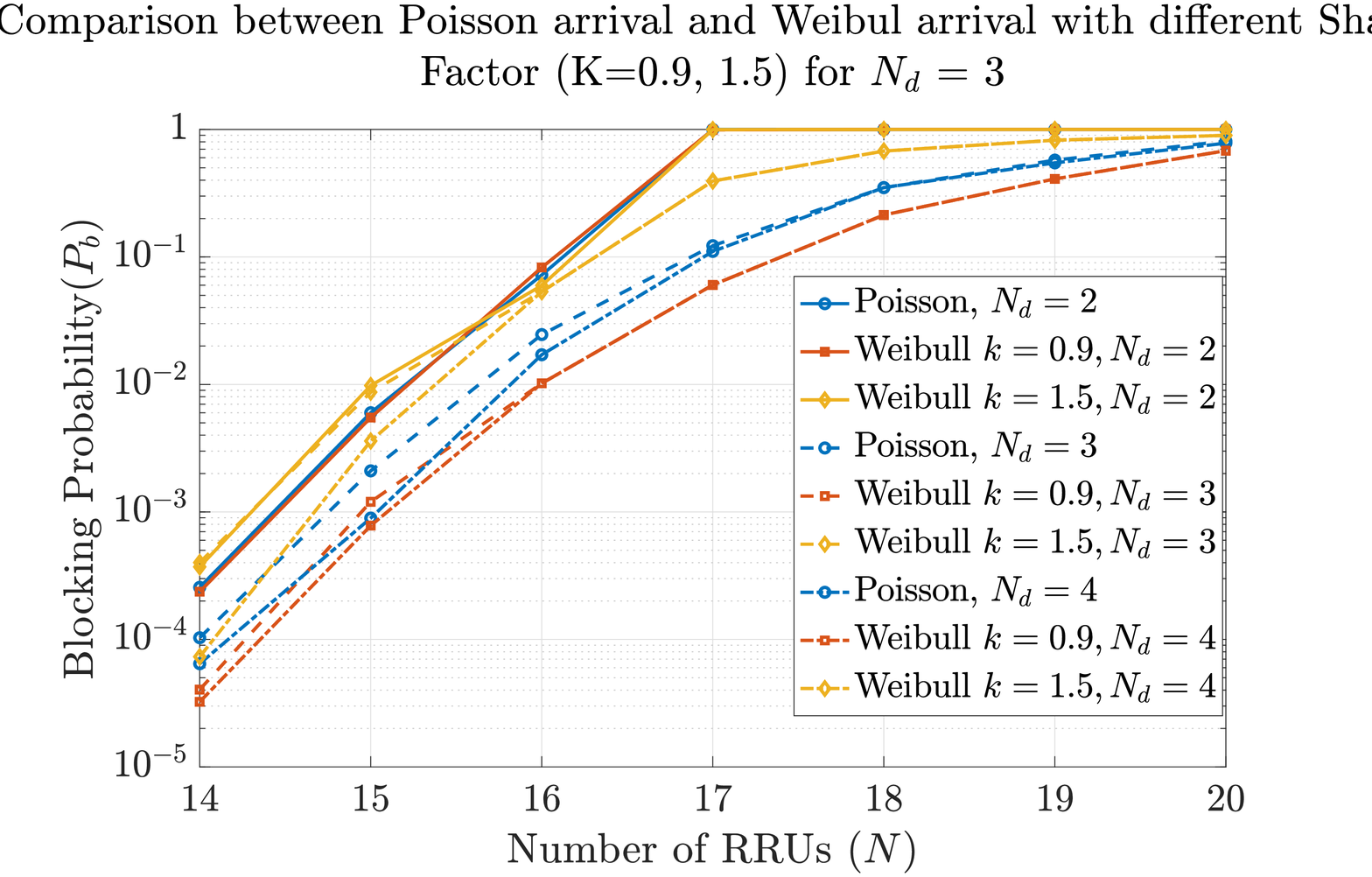}
		\vspace{-0.3in}
		\caption{Performance comparison between Poisson and Weibull Distribution for $a=0.3$, $N_d=2,3,4$, $F_l-R_l=1$}
		\label{fig:Comp_poisson_weibul_a_0.3}
	\end{figure}
    Figure \ref{fig:Comp_poisson_weibul_a_0.3} shows comparative result of Weibull arrival process with two shape factors ($k=0.9$ and $k=1.5$) against already discussed results using Poisson arrivals under the normalized traffic load $a=0.3$. We notice that our results with Poisson arrival process lies between Weibull process with $k=1.5$ and $k=0.9$. The reason for this is that the intensity of the arrival process changes according to the shape factor. For example, if the average inter-arrival time for our original exponential distribution is $1/\lambda$ (or the arrival rate for our original Poisson distribution is $\lambda$), then the inter-arrival time for Weibull  arrival process is $\Gamma (1+1/k)/\lambda = 1.0522/\lambda$  for the shape factor ($k$)=0.9. Therefore, we see that the inter arrival time  gets larger which implies a decrease in arrival rate. Thus a Weibull distribution with $k=0.9$ yields a lower blocking probability compared to our Poisson arrival case. The same argument can be used to explain the reason for  $k=1.5$ having a higher blocking probability when compared to our Poisson arrival case. In addition, as the average rate of arrival increases for higher values of $a$ (which is not shown here), the difference becomes less pronounced as higher capacity will move the RRU states towards the higher bandwidth, thus masking the difference in distribution.

\section{Conclusions} \label{sec:Conclusions}
	In this paper, we have introduced the concept of variable rate fronthaul  for cloud-RANs. After providing the description of the network architecture, we have formulated a mathematical description of our model. We have used Queuing Theory with a two-phase approach to solve the model and obtain an analytical form for blocking probability at the fronthaul aggregator. 
    We have then performed simulation of the model using Matlab's discrete event simulator, and the results were compared with those obtained from our analytical model.
    
    Besides showing a close match between analytical and simulation models, our results prove that by dynamically varying the cell's bandwidth, according to the actual end user demand, we can achieve a more efficient fronthaul transport of a group of C-RAN cells, without increasing the complexity, cost and energy consumption of the RRUs. This is especially relevant for next generation of high-density cell deployment, as multiplexing several cells, can sensibly lower fronthaul costs. 
    
\section*{Acknowledgment} \label{sec:Acknowledgment}
	Financial support from SFI 14/IA/2527 (O'SHARE) and 13/RC/2077 (CONNECT) is gratefully acknowledged.

%
%
%
%
\section*{APPENDIX A} \label{sec:appendixas}
Since $\mathbf{k}$ is reversible, the local balance equation 
\begin{equation}\label{eqn:locBalAggregatorSteadyState}
	\mathop{P}(\mathbf{k}^{(i)}). Q_{\mathbf{k}^{(i)}\mathbf{k}^{(j)}} = \mathop{P}(\mathbf{k}^{(j)}).Q_{\mathbf{k}^{(j)}\mathbf{k}^{(i)}}
\end{equation}
holds at the steady state. Without loss of generality, let 
\[\mathbf{k}^{(i)} = \{ k_1, \dots, k_m, k_{m+1},\dots, k_M\}^T,\]
\[\mathbf{k}^{(j)} = \{ k_1, \dots, k_m-1, k_{m+1}+1,\dots, k_M\}^T\]
substituting (\ref{eqn:aggregatorTransition}) into (\ref{eqn:locBalAggregatorSteadyState}), we get
\begin{equation}\label{eqn:steadystLocBalEqn_1}
	\begin{array}{l} 
		\!\!\!\!\!\!\!\! \mathop{P}(k_1, \dots, k_m, k_{m+1},\dots, k_M)k_m \lambda_m  = \\
		\,\, \mathop{P}(k_1, \dots, k_m\!-\!1, k_{m+1}\!+\!1,\dots, k_M)(k_m\!+\!1) \mu_{m+1}
	\end{array} 
\end{equation}
Expression in (\ref{eqn:steadystLocBalEqn_1}) can be re-written as
\begin{equation}\label{eqn:steadystLocBalEqn_2}
	\frac{\mathop{P}(k_1, \dots, k_m\!\!-\!\!1, k_{m+1}\!+\!1,\dots, k_M)}{\mathop{P}(k_1, \dots, k_m, k_{m+1},\dots, k_M\!)} \!=\!\! \frac{k_m}{(k_m\!+\!1)} \frac{\lambda_m}{\mu_{m+1}} \!
\end{equation}
After little manipulation of (\ref{eqn:steadystLocBalEqn_2}), we obtain  (\ref{eqn:steadystLocBalEqn_3}).
\begin{equation}\label{eqn:steadystLocBalEqn_3}
	\frac{\mathop{P}(k_1, \dots, k_{m-1}, k_m,\dots, k_M)}{\mathop{P}(k_1, \dots, k_{m-1}\!+\!1, k_m\!-\!1,\dots, k_M\!)} \!=\!\! \frac{k_{m-1}\!+\!1}{k_m} \frac{\lambda_{m-1}}{\mu_m} \!
\end{equation}
Clearly, (\ref{eqn:steadystLocBalEqn_3}) is iterative, therefore we can iterate this equation to obtain the following:
\begin{equation}\label{eqn:steadystLocBalEqn_4}
	\begin{array}{l}
		\displaystyle \mathop{P}(k_1, \dots, k_{m-1}, k_m,\dots, k_M) = \\
		\displaystyle \mathop{P}(k_1, \dots, k_{m-1}\!+\!k_m, 0,\dots, k_M\!) \Bigg\{\!\frac{\lambda_{m-1}}{\mu_m} \!\Bigg\}^{\!\!\!k_m}  \!\! \Bigg\{\! \! \frac{(k_{m-1}\!+\!k_m)!}{k_{m-1}! \,\, k_m!} \!\!\Bigg\}
	\end{array}
\end{equation}
Using (\ref{eqn:steadystLocBalEqn_4}), we can write:
\begin{equation}\label{eqn:steadystLocBalEqn_5}
	\begin{array}{l}
		\displaystyle \mathop{P}(k_1, \dots, k_{M-1}, k_M) = \\
		\displaystyle \mathop{P}(k_1, \dots, k_{M-1}\!+\!k_M, 0\!) \Bigg\{\!\frac{\lambda_{M-1}}{\mu_M} \!\Bigg\}^{\!\!\!k_M}  \!\! \Bigg\{\! \! \frac{(k_{M-1}\!+\!k_M)!}{k_{M-1}! \,\, k_M!} \!\!\Bigg\}
	\end{array}
\end{equation}
Then, starting with (\ref{eqn:steadystLocBalEqn_5}) and iterating over all the entries except the first position we obtain:
\begin{equation}\label{eqn:steadystLocBalEqn_6}
	\begin{array}{l}
		\displaystyle \!\!\!\!\!\!\!\!\!\mathop{P}(k_1, k_2, \dots,  k_M) =\\
		\,\, \displaystyle \mathop{P}(K_s, 0, 0 , \dots , 0)\Bigg[\!\! \frac{K_s!}{\prod_{i=1}^{M}k_i!}   \prod_{i=2}^{M} \! \bigg(\!  \frac{\lambda_{i-1}}{\mu_i}  \! \bigg)^{\!\sum_{j=i}^{M}k_j} \Bigg] \!\!\!\!\\
	\end{array}
\end{equation}
In (\ref{eqn:steadystLocBalEqn_6}), $K_s = \sum_{i=1}^{M}k_i$. Now, with the help of (\ref{eqn:aggregatorTransition}), we can write the following flow balance equation for $N \leq N^{\text{max}}_{\text{RRU}}$:
\begin{equation}\label{eqn:steadystLocBalEqn_7}
	\begin{array}{l} 
		\!\!\!\!\!\!\! \mathop{P}(K, 0, \dots, 0) (N \!-\! K)  \lambda  =  \mathop{P}(K\!+\!1, 0, \dots, 0)(K\!+\!1) \mu_{1} \!\!\!\!
	\end{array}
\end{equation}
Expression in (\ref{eqn:steadystLocBalEqn_7}) can be simplified using the same process followed in (\ref{eqn:steadystLocBalEqn_4}) to obtain the following expression:
\begin{equation}\label{eqn:steadystLocBalEqn_8}
	\begin{array}{l} 
		\!\!\!\!\!\!\! \displaystyle \mathop{P}(K, 0, \dots, 0)  =  \binom{N}{K} \bigg(\! \frac{\lambda}{\mu_1}  \!\bigg)^{\!\!\!K} \mathop{P}(0, 0, \dots, 0) \!\!\!\!
	\end{array}
\end{equation}
Substituting (\ref{eqn:steadystLocBalEqn_8}) into (\ref{eqn:steadystLocBalEqn_6}) and after some manipulation, we get the following final deduction:
\begin{equation}\label{eqn:AppxsteadystLocBalEqn_9}
	\begin{array}{l}
		\displaystyle \!\!\mathop{P}(k_1, k_2, \dots,  k_M) =   \vspace{0.3cm} \\
		\displaystyle \quad \mathop{P}(0, 0, \dots, 0)\Bigg[\!\!\binom{N}{K_s} \frac{K_s!}{\prod_{i=1}^{M}k_i!}   \prod_{i=1}^{M} \! \bigg(\!  \frac{\lambda_{i-1}}{\mu_i}  \! \bigg)^{\!\sum_{j=i}^{M}k_j} \Bigg] \!\!\!\!
	\end{array}
\end{equation}\\
Following the similar procedure, as discussed obtain (\ref{eqn:steadystLocBalEqn_7}) to (\ref{eqn:AppxsteadystLocBalEqn_9}), we can obtain the expression for $N > N^{\text{max}}_{\text{RRU}}$ as provided in (\ref{eqn:AppxsteadystLocBalEqn_10})
\begin{equation}\label{eqn:AppxsteadystLocBalEqn_10}
\begin{array}{l}
\displaystyle \!\!\mathop{P}(k_1, k_2, \dots,  k_M) =   \vspace{0.3cm} \\
\displaystyle \quad \mathop{P}(0, 0, \dots, 0)\Bigg[\!\!\binom{\!N^{\text{max}}_{\text{RRU}}\!}{K_s} \frac{K_s!}{\prod_{i=1}^{M}k_i!} \!  \prod_{i=1}^{M} \!\! \bigg(\!\!  \frac{\lambda_{i-1}}{\mu_i}  \!\! \bigg)^{\!\!\sum_{j=i}^{M}k_j} \Bigg] \!\!\!\!
\end{array}
\end{equation}\\

\section*{APPENDIX B} \label{sec:appendixb}
\begin{figure}[h]
		\includegraphics[clip, trim={0.0in, 0, 0.5in, 0in }, width=\linewidth]{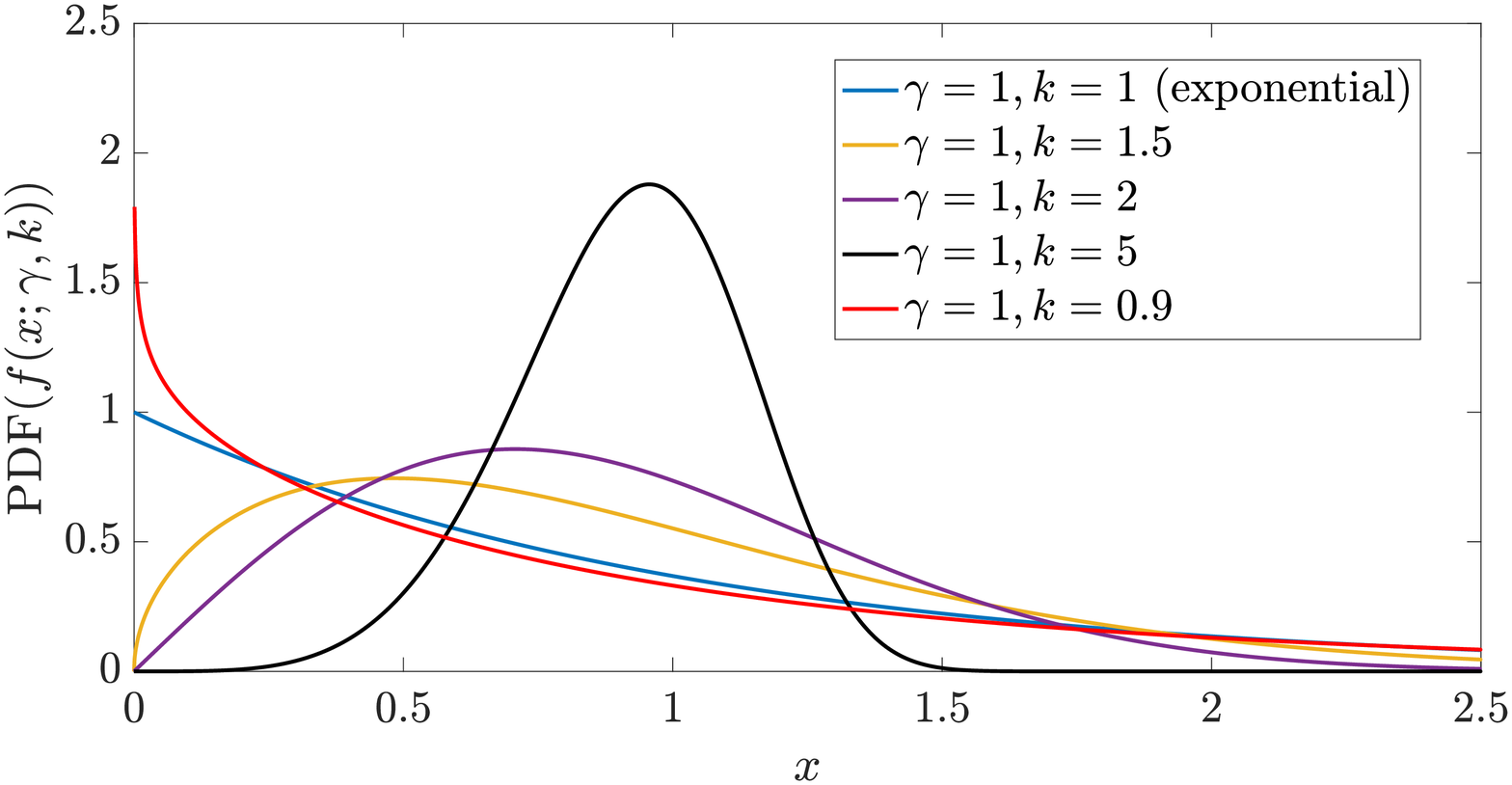}
		\vspace{-0.3in}
		\caption{PDF of Weibull distribution with different shape factors}
		\label{fig:weibullDist}
	\end{figure}

The additional simulations are based on the use of a Weibull distribution, which generalizes a large class of distributions (exponential, Rayleigh, chi-squared etc.) depending on the shape factor $k$. For example, for $k=1$ we obtain an exponential distribution (Poisson arrival process = exponentially distributed inter-arrival times ). Equation-(\ref{eqn:wblpdf}) provides the details of an Weibull distribution with mean = $\gamma \Gamma (1+1/k)$ and variance = $\gamma^2 \left[\Gamma(1+2/k) - (\Gamma(1 + 1/k))^2 \right]$. Figure \ref{fig:weibullDist} plots the distribution under different shape factor ($k$).
	\begin{align}
	\label{eqn:wblpdf}
	f(x,\gamma,k) =
	\begin{cases}
	\frac{k}{\gamma}\left(\frac{x}{\gamma}\right)^{k-1} e^{-(x/\gamma)^k} \qquad \text{for } x\geq0  \\
	0 \qquad \qquad\qquad \qquad \quad \text{for } x<0
	\end{cases} 
	\end{align}
	
\bibliographystyle{./References/ieeetran}
\bibliography{./References/references}

\end{document}